\documentclass[12pt]{article} \hoffset=-0.5in 
\newcommand{\df}{\ {\overset {\rm def} =}\ }
\newcommand{\dr}[2]{\frac {{\rm d} {#1}} {{\rm d} {#2}}}

\newcommand{\dril}[2]{{{\rm d} {#1}} / {{\rm d} {#2}}}

\newcommand{\llim}[1] {\ {\underset {#1} {\longrightarrow}}\ }

 \usepackage[dvips]{graphicx}
 \usepackage{amsmath}
 \usepackage{amsfonts}
 \usepackage{amssymb}

\begin{document}

\title{Can a charged dust ball be sent through the Reissner--Nordstr\"{o}m
wormhole?}

\author{Andrzej Krasi\'nski and Krzysztof Bolejko\thanks{This research was
supported by the Polish Ministry of Education and Science grant no 1 P03B 075
29.}}

\date {}

\maketitle

\centerline{N. Copernicus Astronomical Center, Polish Academy of Sciences}

\centerline{Bartycka 18, 00 716 Warszawa, Poland}

\centerline{email: akr@camk.edu.pl}

\begin{abstract}
In a previous paper we formulated a set of necessary conditions for the
spherically symmetric weakly charged dust to avoid Big Bang/Big Crunch, shell
crossing and permanent central singularities. However, we did not discuss the
properties of the energy density, some of which are surprising and seem not to
have been known up to now. A singularity of infinite energy density does exist
-- it is a point singularity situated on the world line of the center of
symmetry. The condition that no mass shell collapses to $R = 0$ if it had $R >
0$ initially thus turns out to be still insufficient for avoiding a singularity.
Moreover, at the singularity the energy density $\epsilon$ is
direction-dependent: $\epsilon \to - \infty$ when we approach the singular point
along a $t =$ const hypersurface and $\epsilon \to + \infty$ when we approach
that point along the center of symmetry. The appearance of
negative-energy-density regions turns out to be inevitable. We discuss various
aspects of this property of our configuration. We also show that a permanently
pulsating configuration, with the period of pulsation independent of mass, is
possible only if there exists a permanent central singularity.
\end{abstract}

\section{The problem}

In our previous paper \cite{KrBo2006} (Paper I) we formulated a set of necessary
conditions for the spherically symmetric weakly charged dust to avoid Big
Bang/Big Crunch, shell crossing and permanent central singularities. In fact, it
had already been proven by Ori \cite{Ori1990, Ori1991} that weakly charged dust
with no central singularity necessarily evolves toward a shell crossing.
However, Ori assumed that the absolute value of the charge density (in geometric
units) is strictly smaller than the mass density throughout the volume,
including the center. His proof did not include the subcase when the values of
these two densities at the center of symmetry are equal. We discussed precisely
this subcase. We found that shell crossings are still unavoidable when the
energy function $E \geq 0$, but the conditions did not lead to a contradiction
when $E < 0$. We also provided an example of a solution of these conditions,
which was in fact incorrect (see below). However, we did not discuss the
properties of the energy density. While we started to investigate more
properties of the numerical example presented in Paper I, it turned out that
this density was becoming negative in a vicinity of the center of symmetry, for
a brief period around the bounce instant. Then, exact consideration showed that
this is a general problem -- a weakly charged spherically symmetric dust ball
will always have such a negative-energy-density region if the charge and mass
densities become equal in absolute value at the center. This fact is hidden in
the equations in a rather tricky way. In the general case treated by Ori this
problem can be avoided by an appropriate choice of the arbitrary functions.

The explicit example we gave in Paper I had a problem - it did not obey the
condition that the ratio $R/{\cal M}^{1/3}$ must have a finite nonzero limit at
the center of symmetry. For our example, this limit was zero, which means that
there was a permanent central singularity in it (the limit of the mass density
at the center was actually $- \infty$). The permanent infinity is easily cured
by changing $x^{4/3}$ to $x^{5/3}$ in the formulae for $\Gamma$ and $E$. The
whole subsequent reasoning then applies with only little quantitative
changes,\footnote{This error has already been pointed out in an erratum to Paper
I, and a fully corrected text is available from the gr-qc archive, see Ref.
\cite{KrBo2006}.} but a direction-dependent point singularity at the center
necessarily appears at the instant of maximal compression.

In the present paper we fill in the gaps left out by previous investigation. We
show that the energy density of the charged dust must become negative within a
finite time-interval containing the instant of maximal compression. We also show
that the point on the world line of the center of symmetry that is the limit of
the points of maximal compression is a direction-dependent singularity of
infinite energy density. The energy density tends to {\it minus} infinity when
we approach the singular point along the hypersurface of maximal compression,
and to plus infinity when we approach the same point along the center of
symmetry. Thus, the condition that the areal radius $R(t, r)$ never goes down to
zero if it was nonzero initially turns out to be still insufficient for avoiding
singularities. These are the most striking and most important results of the
present paper.

In order to make the paper independently readable, we quote the basic results of
Paper I in the next section, without proof. All the proofs (or references to the
proofs) and details of the calculation can be found in Paper I. Section
\ref{paper1} also contains remarks on the interpretation of the energy density,
which is nontrivial in the presence of charges, and was not properly discussed
in the earlier papers. In section \ref{negdens} we show that the quantity $u =
\Gamma - QQ,_N / R$ must become negative in a vicinity of the instant of maximal
compression whenever the conditions ${Q,_N}^2 = G / c^4$ at the center and
${Q,_N}^2 < G / c^4$ elsewhere are obeyed. In section \ref{negucons} we discuss
the various consequences of this fact and show that this necessarily implies a
negative energy density for a certain period around the bounce instant. In
section \ref{novac} we show that the negative energy density regions cannot be
eliminated by assuming that the charged dust ball contains an empty (Minkowski)
vacuole around the center of symmetry because such a vacuole cannot exist. In
section \ref{transient} we prove the existence of the point-singularity on the
world line of the center of symmetry. In section \ref{perm} we show that, even
with negative energy density allowed, a solution with the period of oscillation
independent of mass does not exist. A permanently nonsingular configuration of
nonstatic weakly charged dust is thus impossible, unless one allows a permanent
central singularity. We also compare our conclusion with the properties of the
uncharged case. Section \ref{conclu} summarizes the conclusions.

\section{Basic formulae and results of Paper I}\label{paper1}

For a spherically symmetric spacetime in comoving coordinates, the metric can be
put in the form
\begin{equation}\label{2.1}
{\rm d} s^2 = {\rm e}^{C(t,r)} {\rm d} t^2 - {\rm e}^{A(t, r)} {\rm d} r^2 -
R^2(t,r)\left[{\rm d} \vartheta^2 + \sin^2(\vartheta) {\rm d} \varphi^2\right].
\end{equation}
In the generic case $R,_r \neq 0$, assuming there are no magnetic charges, the
Einstein--Maxwell equations yield the following result \cite{KrBo2006, PlKr2006,
Vick1973}. The only independent nonzero component of the electromagnetic field
is
\begin{eqnarray}
F^{01} &=& Q(r) {\rm e}^{- (A + C)/2} / R^2, \label{2.2} \\
Q,_r &=& (4\pi / c) \rho_e {\rm e}^{A/2} R^2, \label{2.3}
\end{eqnarray}
where $Q(r)$ is an arbitrary function -- the electric charge within the
$r$-surface, and $\rho_e$ is the electric charge density;
\begin{equation}\label{2.4}
\frac {\kappa} 2 \epsilon R^2 {\rm e}^{A/2} = \frac G {c^4}\ N,_r, \qquad \kappa
\df 8 \pi G/c^4,
\end{equation}
where $\epsilon$ is the energy density and $N,_r$ is an arbitrary function of
integration. The $N$ so defined corresponds, in the electrically neutral case,
to the energy equivalent to the sum of rest masses within the $r$-surface. With
charges present, this interpretation involves a subtle point, see below after
eq. (\ref{2.10}). The ratio $Q,_r/N,_r = \rho_e/(c \epsilon)$ is
time-independent.
\begin{equation}\label{2.5}
{\rm e}^{A/2} = \frac {R,_r} {\Gamma(r) - QQ,_N / R},
\end{equation}
where $Q,_N$ is just an abbreviation for $Q,_r/N,_r$;
\begin{equation}\label{2.6}
C,_r = 2 \frac {{\rm e}^{A/2}} {R^2} QQ,_N;
\end{equation}
\begin{equation}\label{2.7}
{\rm e}^{- C}{R,_t}^2 = \Gamma^2 - 1 + \frac {2M(r)} R + \frac
{Q^2\left({Q,_N}^2 - G/c^4\right)} {R^2} - \frac 1 3 \Lambda R^2,
\end{equation}
where $\Lambda$ is the cosmological constant, and $M(r)$ is an arbitrary
function. By analogy with the uncharged Lema\^{\i}tre -- Tolman model,
$(\Gamma^2 - 1)/2$ is often denoted as $E(r)$.

The function $M(r)$ is the {\it effective mass}, and it {\it need not be
positive}. It is connected with the previously defined arbitrary functions by
\begin{equation}\label{2.8}
\frac G {c^4} \Gamma N,_r = \left(M + QQ,_N\Gamma\right),_r.
\end{equation}
The quantity
\begin{equation}\label{2.9}
{\cal M} \df M + QQ,_N\Gamma,
\end{equation}
is the active gravitational mass. Thus, via (\ref{2.8}), $\Gamma$ determines by
how much ${\cal M}$ increases when a unit of rest mass is added to the source,
i.e. $\Gamma$ is a measure of the gravitational mass defect/excess.

The energy density of the dust is:
\begin{equation}\label{2.10}
\kappa \epsilon = \frac {2GN,_r} {c^4 R^2 R,_r}\ \left(\Gamma - \frac {QQ,_N}
R\right).
\end{equation}
Now note the subtle point. In the electrically neutral case, $Q = 0$, the
quantity $\epsilon$ is the density of rest-mass (in energy units), so that the
3-space integral $\int_V \sqrt{- g} \epsilon {\rm d}_3x$ simply equals the sum
of rest masses in the volume $V$ divided by $c^2$. However, with charges
present, $\epsilon$ contains a contribution from the charges. As will be shown
later in this paper, $QQ,_N \equiv \frac 1 2 \left(Q^2\right),_r / N,_r$ must be
positive in a vicinity of the center of symmetry, and $R > 0$ everywhere for
geometrical reasons. Thus, the presence of charges always {\it decreases} the
energy density. It will turn out later that, in the class of models considered
here, $\epsilon$ necessarily becomes negative for a brief period around the
instant of maximal compression along each world-line except the central one.

The functions $R(t, r)$ and $C(t, r)$ are implicitly defined by the set
(\ref{2.6}) -- (\ref{2.7}), which can in general be solved only numerically.

If the configuration considered here is matched to the Reissner -- Nordstr\"{o}m
(R--N) metric across a hypersurface $r = r_b$, then the following must hold
\cite{PlKr2006, Vick1973, KrBo2006}:
\begin{equation}\label{2.11}
e = \frac {\sqrt{G}} {c^2} Q(r_b), \qquad m = \left(M + QQ,_N \Gamma\right)_{r =
r_b},
\end{equation}
where $e$ and $m$ are the R--N charge and mass parameters, respectively.

Unlike in the electrically neutral case, in charged dust the Big Bang/Big Crunch
(BB/BC) singularity can be avoided, i.e. there exist solutions of (\ref{2.6}) --
(\ref{2.7}) in which the function $R(t, r)$ never goes down to zero if it was
nonzero initially.\footnote{$R(t, r) = 0$ permanently at the center of
symmetry.} We give the conditions for the existence of such solutions only in
the case $\Lambda = 0$. One of those conditions is
\begin{equation}\label{2.12}
M^2 \geq 2EQ^2\left({Q,_N}^2 - G/c^4\right),
\end{equation}
which is fulfilled identically when $E = 0$. With (\ref{2.12}) fulfilled and $E
\neq 0$, the right-hand side of (\ref{2.7}) has two roots, given by
\begin{equation}\label{2.13}
R_{\pm} = - \frac M {2E} \pm \frac 1 {2E} \sqrt{M^2 - 2EQ^2\left({Q,_N}^2 -
G/c^4\right)}.
\end{equation}
The second condition for avoiding the BB/BC singularity depends on $E$:

(a) When $E \leq 0$, the condition is
\begin{equation}\label{2.14}
{Q,_N}^2 < G/c^4 \qquad {\rm and}\qquad M > 0.
\end{equation}
With (\ref{2.14}) fulfilled and $E < 0$, $R$ oscillates between the nonzero
minimum at $R = R_+$ and the maximum at $R = R_-$. With (\ref{2.14}) and $E =
0$, $R$ goes down from infinity to the finite minimal value $R_{\rm min} = Q^2
\left(G/c^4 - {Q,_N}^2\right) / (2M)$ and then increases to infinity again.

(b) When $E > 0$ and $M > 0$, the BB/BC singularity does not occur if ${Q,_N}^2
< G/c^4$. When $E > 0$ and $M < 0$, nonsingular solutions exist with no further
conditions, provided $R > R_+$ initially. The bounce with $M < 0$ is
nonrelativistic, since it occurs also in Newton's theory, under the same
conditions ($M < 0$ means that the electric repulsion of the charges spread
throughout the volume of the dust prevails over the gravitational attraction of
the mass).

The inequality ${Q,_N}^2 < G / c^4$ translates into $\left|\rho_e\right| <
\sqrt{G} \epsilon / c$, which means that, in geometric units, the absolute value
of the charge density is {\it smaller} than the mass density.\footnote{This
includes also the case $Q,_N = 0$, i.e. zero charge density, provided $Q \neq
0$, i.e. nonzero total charge. Such a configuration is neutral dust moving in
the exterior electric field of a spherically symmetric source. Also in this
case, the BB/BC singularity is avoided.}

There is another subtle point here. The conditions (\ref{2.14}) guarantee that a
particle that had $R > 0$ initially will not hit the set $R = 0$ in the future
or in the past. But, as we will see in Sec. \ref{transient}, the configurations
obeying (\ref{2.14}) contain a cleverly hidden singularity of a type hitherto
unknown in dust solutions. On the world line of the center of symmetry, where
$R(t, r) = 0$ permanently, there is a point in which $\epsilon \to + \infty$ for
a single instant. This instant is the limit at $R \to 0$ of the hypersurface
$S_{\rm min}$ consisting of the instants in which the mass shells with $R > 0$
attain their minimal sizes. However, if we approach the same spacetime location
along the hypersurface $S_{\rm min}$, then $\epsilon \to - \infty$.

The surface of the charged sphere obeys the equation of radial motion of a
charged particle in the Reissner--Nordstr\"{o}m spacetime. Thus, the surface of
a collapsing sphere must continue to collapse until it crosses the inner R--N
horizon $r_-$, and can bounce at $R \leq r_-$. Then, however, it cannot
re-expand back into the same spacetime region from which it collapsed, as this
would require motion backward in time. The surface would thus continue through
the tunnel between the singularities and re-expand into another copy of the
asymptotically flat region.

At small charge density (${Q,_N}^2 < G / c^4$ throughout the volume) a shell
crossing is unavoidable, and it will block the passage through the R--N tunnel,
as shown by Ori \cite{Ori1990, Ori1991}. A nonsingular bounce might be possible
only if ${Q,_N}^2 > G / c^4$ everywhere or if ${Q,_N}^2 \to G / c^4$ at the
center, while ${Q,_N}^2 < G / c^4$ elsewhere. The first case was dealt with by
Ori (unpublished \cite{Oripriv}). We consider the second case in Section 3, and
the result is that a nonsingular bounce might happen only if the energy density
becomes negative for a period around the bounce instant.

Shell crossings are most conveniently discussed in the mass-curvature
coordinates $({\cal M}, R)$, first introduced by Ori \cite{Ori1990}. Details of
the transformation can be found in Ref. \cite{PlKr2006}. The $({\cal M}, R)$
coordinates allow the Einstein--Maxwell equations with $\Lambda = 0$ to be
explicitly integrated, but it happens at a price. The spacetime points are
identified by the values of ${\cal M}$ and $R$, which means a given point is
defined by saying ``it is the place where the shell containing the mass ${\cal
M}$ has the radius $R$''. The information on the time-dependence of $R$ is lost,
and can be regained only by reverting to the comoving coordinates -- but the
transformation equations are equivalent to (\ref{2.6}) -- (\ref{2.7}) and cannot
be explicitly integrated.

The solution of the Einstein -- Maxwell equations is given below. The velocity
field has only one contravariant component:
\begin{equation}\label{2.15}
u^R = \pm \sqrt{\Gamma^2 - 1 + \frac {2M} R + \frac {Q^2\left({Q,_N}^2 -
G/c^4\right)} {R^2} - \frac 1 3 \Lambda R^2}
\end{equation}
($+$ for expansion, $-$ for collapse). We define the auxiliary quantities
\begin{equation}\label{2.16}
u \df \Gamma - QQ,_N/R, \qquad \Delta \df 1 - \frac {2 {\cal M}} R + \frac
{GQ^2} {c^4 R^2} + \frac 1 3 \Lambda R^2,
\end{equation}
and we get for the metric
\begin{equation}\label{2.17}
g_{{\cal M} {\cal M}} = F^2 \Delta,  \qquad g_{{\cal M} R} = F u / u^R, \qquad
g_{R R} = 1 / (u^R)^2,
\end{equation}
where the function $F({\cal M}, R)$ is given by
\begin{equation}\label{2.18}
F,_R = - \frac 1 {\left(u^R({\cal M}, R)\right)^3} \left\{\Gamma,_{\cal M} +
\frac 1 {R \Gamma}\left[1 - \frac {c^4} G \left({Q,_N}^2 +
QQ,_{NN}\right)\right]\right\}.
\end{equation}
With $\Lambda = 0$, the integral of this is elementary, but to give its explicit
form several cases have to be considered separately (see Ref \cite{Ori1990} for
a list). Shell crossings occur at the zeros of the function $F$. Thus, to avoid
shell crossings, the arbitrary functions must be chosen so that $F \neq 0$
everywhere.

The only nonvanishing components of the electromagnetic tensor in the $({\cal
M}, R)$ coordinates are
\begin{equation}\label{2.19}
F^{{\cal M} R} = - F^{R {\cal M}} = \frac Q {FR^2}, \qquad F_{{\cal M} R} = -
F_{R {\cal M}} = - \frac {FQ} {R^2},
\end{equation}
while the charge density and the energy-density are
\begin{equation}\label{2.20}
\frac {4\pi \rho_e} c = - \frac {Q,_{\cal M}} {R^2 F u^R}, \qquad \kappa
\epsilon = - \frac 2 {\Gamma R^2 F u^R}.
\end{equation}

The $({\cal M}, R)$-coordinates cover only such a region where $R,_t$ has a
constant sign. The function $F$ changes sign where $R,_t$ does, and so does $u^R
= {\rm e}^{- C/2} R,_t$. Thus, $Fu^R$ preserves its sign when collapse turns to
expansion and vice versa.

In the $({\cal M}, R)$ coordinates eq. (\ref{2.8}) reads
\begin{equation}\label{2.21}
{\cal M},_N = (G \Gamma / c^4).
\end{equation}

Just as in the L--T model, the set $R = 0$ in charged dust consists of the Big
Bang/Crunch singularity (which is now avoidable), and of the center of symmetry,
which may or may not be singular. The conditions for the absence of a permanent
central singularity are the following (not all of them are independent, but this
is the full list):
\begin{eqnarray}
N(r_c) &=& 0 = Q(r_c) = {\cal M}\left(r_c\right) = M\left(r_c\right),
\label{2.22} \\
\lim_{r \to r_c} R/{\cal M}^{1/3} &=& \beta(t) \neq 0, \label{2.23} \\
\lim_{r \to r_c} \Gamma^2(r) &=& 1 \Longrightarrow \lim_{r \to r_c} E(r) = 0,
\label{2.24} \\
\lim_{r \to r_c} 2E / {\cal M}^{2/3} &=& \lim_{r \to r_c} \left(\Gamma^2(r) -
1\right) / {\cal M}^{2/3} \df D = {\rm const}, \label{2.25}
\end{eqnarray}
and this last constant may be zero.

In the mass-curvature coordinates, with $\Lambda = 0$, it is easy to solve the
evolution equation $\dril R s = u^R$. We quote here the only one case, for which
we found in Ref. \cite{KrBo2006} that there may exist solutions avoiding both
kinds of singularity, $E < 0$ (see Fig. 5 in Paper I).

We take $\ell = +1$ for expansion and $\ell = -1$ for collapse at the initial
instant of evolution, and we denote
\begin{equation}\label{2.26}
\Phi \df \frac {Q^2 \left({Q,_N}^2 - G / c^4\right)} {2E}.
\end{equation}
When $E < 0$, a solution exists only with $\Phi < M^2 / \left(4E^2\right)$, and
it is given by the parametric equations
\begin{eqnarray}\label{2.27}
R &=& - \frac M {2E} - \sqrt{\frac {M^2} {4 E^2} - \Phi}\ \cos \omega, \nonumber
\\
s - s_B({\cal M}) &=& \frac {\ell} {\sqrt{- 2E}}\ \left(- \frac M {2E}\ \omega -
\sqrt{\frac {M^2} {4 E^2} - \Phi}\ \sin \omega\right),
\end{eqnarray}
where $\omega$ is a parameter and $s_B({\cal M})$ is an arbitrary function of
integration. This will avoid a BB/BC singularity only if $M > 0$ and $\Phi
> 0$. At $s = s_B$, $R$ starts off with the minimal value $R_{\rm min} = R_+$
(see eq. (\ref{2.13})), and periodically returns to this value, never going down
to zero if $R_{\rm min} \neq 0$. The period is
\begin{equation}\label{2.28}
T_p = 2 \pi M / (- 2E)^{3/2}.
\end{equation}

The period given by (\ref{2.28}) is with respect to the proper time of the given
shell of constant ${\cal M}$. To calculate the period in the time coordinate
$t$, we must first transform the variables in eq. (\ref{2.7}), to make $({\cal
M}, R)$ the independent variables and $t$ the unknown function. We obtain
\begin{equation}\label{2.29}
T_c = 2 \int_{R_+}^{R_-} \frac {{\rm e}^{-C({\cal M}, R)/2} {\rm d} R}
{\sqrt{\Gamma^2 - 1 + 2M/R + Q^2 \left({Q,_N}^2 - G / c^4\right) / R^2 - (1/3)
\Lambda R^2}}.
\end{equation}
Note that each of the periods, $T_p$ and $T_c$, is a function of ${\cal M}$
only. We found in Paper I that even for a single bounce to be free of shell
crossings, it is necessary that the bounce occurs in a time-symmetric manner,
i.e. all the mass shells have to go through their minimal sizes at the same
instant of the coordinate time $t$.

Finally, we recall the 9 necessary conditions that the functions defining the
model must obey in order that even a single nonsingular bounce can occur. We
denote
\begin{equation}\label{2.30}
F_1 \df 1 - \left(c^4 / G\right) \left({Q,_N}^2 + QQ,_{NN}\right),
\end{equation}
and the conditions are:

(1) $E < 0$;

(2) $E \geq -1/2$;

(3) $\lim_{r \to r_c} F_1/{\cal M}^{1/3} = 0$;

(4) $\Gamma,_{\cal M} < 0$;

(5) ${Q,_N}^2 < G / c^4$ at $N > 0$ and ${Q,_N}^2 = G / c^4$ at $N = 0$;

(6) $M \equiv {\cal M} - QQ,_N \Gamma > 0$;

(7) $M^2 - 2E Q^2 \left({Q,_N}^2 - G / c^4\right) > 0$;

(8)
\begin{equation}\label{2.31}
F_1 > \frac {\Gamma \Gamma,_{\cal M}} {2E}\ M.
\end{equation}

(9)
\begin{equation}\label{2.32}
F_1 > \frac {\Gamma \Gamma,_{\cal M}} {2E}\ \left[M + \sqrt{M^2 - 2EQ^2
\left({Q,_N}^2 - G / c^4\right)}\right].
\end{equation}
Conditions (1) -- (9) must hold in a {\it neighbourhood} of the center. {\it At}
the center, the left-hand sides of conditions (1) and (6 -- 9) must have zero
limits. In addition to this, all the regularity conditions at the center must be
obeyed. Note that the conditions (\ref{2.22}) and (\ref{2.24}), together with
eq. (\ref{2.21}), imply that
\begin{equation}\label{2.33}
\lim_{N \to 0} M/N = 0
\end{equation}
-- this will prove useful later.

For practical calculations, the most convenient radial coordinate is $N(r)$, and
it will be used in some of the following sections.

In Paper I we gave an example of a configuration that was supposed to obey
conditions 1 -- 9 and all the regularity conditions. In fact, it did not obey
the condition (\ref{2.23}). This oversight has already been corrected in an
erratum, and in the version stored in the gr-qc archive, see Ref.
\cite{KrBo2006}. That paper contained one more error: in the caption to Fig. 10
we stated that the two curves that seemed tangent were merely adjacent to each
other. In truth, they {\it were} tangent -- the minimal radius at bounce must be
equal to the radius of the inner R--N horizon at every local extremum of the
latter. The proof is given here in Appendix \ref{extrem}. This is just a
correction of an error that has no relation to the other results of the present
paper.

\section{Inevitability of negative values of $u$}\label{negdens}

\setcounter{equation}{0}

From now on, we assume $\Lambda = 0$.

From eq. (\ref{2.10}) we see that if the quantity $u$ defined in (\ref{2.16})
changes sign at a certain point $(t, r)$, while $R,_r$ does not, then the energy
density also changes sign at that point. We will discuss the consequences of the
change of sign of $u$ in the next section. In the present section we will show
that, with the 9 conditions listed in Sec. 2 fulfilled, $u$ is necessarily
negative within an interval containing the instant of maximal compression
(minimal size) of every mass shell on which ${\cal M} \neq 0$. We first prove
that negative $u$ appears in the hypersurface of maximal compression, and later
we will show that $u / R,_r < 0$ for a certain period around the instant of
maximal compression at every ${\cal M} \neq 0$.

At first sight, it seems that the conditions (1) -- (9) given at the end of Sec.
2 secure $u > 0$ in a vicinity of the center, since $\Gamma \to 1$, $Q/N \to \pm
\sqrt{G}/c^2$ and $R/N^{1/3} \to \beta > 0$ as $r \to r_c$. However, the
situation changes if $R = R_+$, i.e. if one approaches the center along the
locus of the inner turning points. Fig. 12 in Ref. \cite{KrBo2006} (and also
Fig. \ref{drawcycles2} further here) shows that along this locus $R$ tends to
zero faster than along any ${\cal M} =$ const line, and eq. (\ref{2.13})
confirms this. At $R = R_+$ we have
\begin{equation}\label{3.1}
u = \Gamma + \frac {Q,_N} {Q \left({Q,_N}^2 - G/c^4\right)} \left[M + \sqrt{M^2
- 2EQ^2 \left({Q,_N}^2 - G/c^4\right)}\right].
\end{equation}
To guarantee $\epsilon > 0$, this should be positive everywhere, including the
center $N = 0$. This requirement can be easily fulfilled if ${Q,_N}^2 < G/c^4$
everywhere including the center. Then, it is enough to choose such $Q(N)$ that
the coefficient in front of the square brackets in (\ref{3.1}) is small and
tends to zero as $N \to 0$, for example $Q(N) = \pm (\sqrt{G}N_0/c^2) {\rm e}^{-
a x^2}$, where $a > 0$ is a constant and $x \df N/N_0$.

In our case, when $Q,_N$ must obey condition (5), it turns out that the limit of
$u$ at $N \to 0$ is necessarily negative. This is seen as follows. We first
observe that
\begin{equation}\label{3.2}
\lim_{N \to 0} \left[- 2EQ^2 \left({Q,_N}^2 - G/c^4\right)\right] / M^2 = 0,
\end{equation}
i.e. that the second term under the square root in (\ref{3.1}) can be neglected
compared to $M^2$ in the limit $N \to 0$ (see Appendix \ref{negli}). Then, from
(\ref{3.1})
\begin{equation}\label{3.3}
\lim_{N \to 0} u = 1 + \lim_{N \to 0} \frac {2MQ,_N} {Q \left({Q,_N}^2 -
G/c^4\right)}.
\end{equation}
Since, by condition (5), $Q,_N(0) \neq 0$, we find
\begin{eqnarray}\label{3.4}
\lim_{N \to 0} u &=& 1 + 2 Q,_N(0) \lim_{N \to 0} \frac {{\cal M} - QQ,_N
\Gamma} {Q \left({Q,_N}^2 - G/c^4\right)} \nonumber \\
  &=& 1 + 2 Q,_N(0) \lim_{N \to 0} \frac {G\Gamma/c^4 - {Q,_N}^2 \Gamma -
QQ,_{NN} \Gamma - QQ,_N \Gamma,_N} {Q,_N \left({Q,_N}^2 - G/c^4\right) +
2QQ,_N Q,_{NN}} \nonumber \\
  &=& -1 + 2 \lim_{N \to 0} \frac {QQ,_{NN} \Gamma - QQ,_N \Gamma,_N}
{{Q,_N}^2 - G/c^4 + 2QQ,_{NN}} \nonumber \\
  &\equiv& -1 + 2 \lim_{N \to 0} \frac {Q,_{NN} \Gamma - Q,_N \Gamma,_N}
{\left({Q,_N}^2 - G/c^4\right)/Q + 2Q,_{NN}} = -1 + \frac {Q,_{NN}(0) - Q,_N(0)
\Gamma,_N(0)} {2Q,_{NN}(0)} \nonumber \\
   &\equiv& - \frac 1 2 - \frac {Q,_N(0) \Gamma,_N(0)} {2Q,_{NN}(0)}.
\end{eqnarray}
(In deriving this, we first applied the de l'Hopital rule, then did an algebraic
simplification in the result, and again applied the de l'Hopital rule in the
denominator.) Now, by condition (5), if $Q,_N(0) > 0$, then $Q,_{NN}(0) \leq 0$,
and if $Q,_N(0) < 0$, then $Q,_{NN}(0) \geq 0$. By condition (4), $\Gamma,_N(0)
\leq 0$. Thus, if $\Gamma,_N(0) < 0 \neq Q,_{NN}(0)$, then the second term in
the last line of (\ref{3.4}) can never be positive. If $\Gamma,_N(0) < 0 =
Q,_{NN}(0)$, then the limit in (\ref{3.4}) is $- \infty$ because $Q,_{NN} < 0$
in a vicinity of $N = 0$. If $\Gamma,_N(0) = 0 \neq Q,_{NN}(0)$, then the limit
is $- 1/2 < 0$. If $\Gamma,_N(0) = 0 = Q,_{NN}(0)$, then the limit of $Q,_N
\Gamma,_N / Q,_{NN}$ at $N \to 0$ is still negative because of conditions (4)
and (5). Consequently, $\lim_{N \to 0} u < 0$ in every case, $\square$.

Since the inequality is sharp (actually, $\lim_{N \to 0} u \leq -1/2$), $u$ will
be negative already at some $N > 0$.

If we approach the center of symmetry along the locus of the outer turning
points, $R = R_-$, then $u$ remains positive up to the very center (see the
proof in the final part of Appendix \ref{negli}). This shows that the
negative-$u$ region does not exist permanently, but appears only for finite time
intervals around the bounce instant.

Thus, a nonsingular bounce of spherically symmetric charged dust is possible
only when there exists a region of negative $u$ for a finite time-interval
around the bounce instant. We discuss the implications of this fact in the next
section.

\section{Consequences of $u < 0$}\label{negucons}

\setcounter{equation}{0}

The only place in the metric where $u$ explicitly appears is via (\ref{2.5}),
which shows that ${\rm e}^A = \left(R,_r / u\right)^2$. This is insensitive to
the sign of $u$. One might thus suspect that it is enough to take ${\rm e}^{A/2}
= - R,_r / u$ instead of (\ref{2.5}) to cure the problem. However, with such
changed ${\rm e}^{A/2}$, the Einstein equations imply that $C,_r = - 2QQ,_N {\rm
e}^{A/2} / R^2$ instead of (\ref{2.6}) (see the derivations in Refs.
\cite{PlKr2006} and \cite{KrBo2006}), i.e. $C$ is sensitive to the sign of $u$.
The changes propagate through all the equations, and the resulting formula for
energy-density, eq. (\ref{2.10}), remains unchanged. This means that $\epsilon$
becomes negative when $u < 0$, unless this change of sign of $u$ can be offset
by a change of sign of $R,_r$ or $N,_r$. We show below that such an offset is
impossible.

The radial coordinate $r$ is arbitrary, and all the formulae are covariant under
the transformation $r = f(r')$, where $f$ is an arbitrary function. The values
of $r$ label flow-lines of the charged dust, all flow lines of the same $r$ form
a 3-cylinder in spacetime, whose sections of constant $t$ are 2-spheres. Thus,
for logical clarity, it is convenient to assume that the labeling is such that
increasing $r$ corresponds to receding from the center of
symmetry.\footnote{This order of labels is reversed on the other side of a shell
crossing, but we are considering models without shell crossings.} Suppose that
$r = r_c$ corresponds to the center of symmetry. Then, $R,_r < 0$ is impossible
in a vicinity of $r = r_c$: since $R = 0$ at $r = r_c$, $R,_r < 0$ at $r > r_c$
would imply $R < 0$ in a vicinity of the center, which is a geometrical
nonsense. Thus, $R,_r > 0$ at the center, and so $R,_r / u < 0$. But then the
energy density will be negative at the center, unless $N,_r < 0$. However, the
physical interpretation of eq. (\ref{2.11}) suggests that $N,_r$ cannot be
negative: since $N = 0$ at the center, $N,_r < 0$ would imply $N < 0$ around the
center. The final conclusion is that the energy density $\epsilon / c^2$ must be
negative where $u < 0$, at least in some neighbourhood of the center (but see
further in this section).

Is the set $u = 0$ a singularity? The answer to this question depends not only
on $u$, but also on the behaviour of $R,_r$. If $R,_r = 0$ at $u \neq 0$, then
this is a shell crossing, which is a curvature singularity. The 9 conditions
listed at the end of section \ref{paper1} were derived from the requirement that
$R,_r \neq 0$ (actually, $R,_r/u \neq 0$) throughout the evolution. All of them
are necessary conditions for $R,_r/u \neq 0$. A sufficient condition is not
known, but anyway we wish to avoid $R,_r/u = 0$, and will not consider this case
here.

If $u/R,_r \neq 0$ at $u = 0$, then there is no singularity at this location,
but the energy density will be negative where $u/R,_r < 0$. We will now
investigate the behaviour of $\epsilon$ in the set $S_{\rm min}$, which is a
3-space composed of those points where $R(t,r)$ assumes its minimal value $R_+$
given by (\ref{2.13}). At those values, $R,_t = 0$.

We have already shown that $u/R,_r < 0$ (implying $\epsilon < 0$) in a vicinity
of the center $R = 0$ in $S_{\rm min}$. Where does $S_{\rm min}$ intersect $u =
0$? At the intersection, from (\ref{2.13}) and (\ref{2.16}), $R$ must obey two
equations:
\begin{eqnarray}\label{4.1}
R = R_+ &=& - \frac 1 {2E} \left[M - \sqrt{M^2 - 2EQ^2 \left({Q,_N}^2 - G /
c^4\right)}\right] \nonumber \\
 &=& \frac {QQ,_N} {\Gamma}.
\end{eqnarray}
We substitute for $M$ from (\ref{2.9}), recall that $\Gamma^2 = 2E + 1$, and
solve (\ref{4.1}) for $Q,_N$:
\begin{eqnarray}\label{4.2}
Q,_N &=& \frac {\Gamma} Q \left({\cal M} - \sqrt{{\cal M}^2 - G Q^2 /
c^4}\right) \nonumber \\
\Longrightarrow R = R_i &=& {\cal M} - \sqrt{{\cal M}^2 - G Q^2 / c^4}.
\end{eqnarray}
This $R_i$ is equal to the radius of the inner Reissner -- Nordstr\"{o}m event
horizon corresponding to the mass shell ${\cal M}$.\footnote{Formally, the
solution of (\ref{4.1}) includes also the outer R--N horizon, with ``+'' in
front of the square root. However, here we consider eq. (\ref{4.1}) within the
set $S_{\rm min}$ that consists of the inner turning points of the various mass
shells. Within this set, the equation $R = {\cal M} + \sqrt{{\cal M}^2 - G Q^2 /
c^4}$ has no solutions, as is known from the general properties of the R--N
solution \cite{PlKr2006}.}

Equation (\ref{4.2}) should be understood as follows. The functions ${\cal
M}(r)$ and $Q(r)$ are arbitrary functions in the model, and $\Gamma$ is
determined by ${\cal M}$ via (\ref{2.8}). Thus, (\ref{4.2}) is an additional
condition imposed on functions that are otherwise arbitrary, and the first of
(\ref{4.2}) may possibly have no solution. However, {\it if} it has a solution,
then this point must have the $R$ value given by the second of (\ref{4.2}).

Then we calculate $R,_r$ at the point determined by (\ref{4.2}), and obtain,
using (\ref{2.8}) and the first of (\ref{4.2}):
\begin{equation}\label{4.3}
\left(R,_r\right)_{R = R_i} = 0.
\end{equation}
Thus, within the set $S_{\rm min}$ the locus of $u = 0$ (if it exists at all)
coincides with the locus of $R,_r = 0$. However, the equations do not allow us
to determine the sign of $u/R,_r$ at this location in the general case. In the
explicit example given in Sec. 8, $u/R,_r < 0$ throughout $S_{\rm min}$, but
this may be a property of that particular model only. In general, we can only
say that there exists a finite neighbourhood of the center $R = 0$ within
$S_{\rm min}$ in which $u / R,_r < 0$.

Analogously, we could consider the set $S_{\rm max}$ consisting of those points
where $R = R_-$, i.e. where all the mass shells attain their maximal radii. We
showed in Appendix \ref{negli} that $u > 0$ in a vicinity of the center within
$S_{\rm max}$. At the center, $R,_r > 0$, so $u / R,_r > 0$ in a neighbourhood
of the center. The whole calculation above can now be repeated with $R_-$
substituted for $R_+$, and the result will be that also within $S_{\rm max}$ the
set $u = 0$ coincides with $R,_r = 0$, but this time at the value $R = R_o =
{\cal M} + \sqrt{{\cal M}^2 - G Q^2 / c^4}$. This means that there exists a
finite neighbourhood of the center $R = 0$ within $S_{\rm max}$ in which $u /
R,_r > 0$, implying $\epsilon > 0$ in the same neighbourhood.

All this implies that a volume of charged dust evolves from negative energy
density in $S_{\rm min}$ to positive energy density in $S_{\rm max}$.
Consequently, the regions of $\epsilon < 0$ and $\epsilon > 0$ are not separated
by a comoving hypersurface -- the matter particles proceed from one to the
other. This should indicate that strong electric fields induce a hitherto
unknown physical process inside matter.

This concludes the discussion of the case when $u / R,_r \neq 0$ at $u = 0$.

Since we have found that $u / R,_r < 0$ in a neighbourhood of $R = 0$ in the
hypersuface $R = R_+$ and $u / R,_r > 0$ in a neighbourhood of $R = 0$ in the
hypersurface $R = R_-$, there must be some intermediate set between them on
which $u / R,_r = 0 = \epsilon$. This set must be a subset of the $u = 0$ set.
Note that in a singularity-free model $R,_r$ can vanish only where $u =
0$,\footnote{Because $R,_r = 0 \neq u$ would be a shell crossing that was
excluded by Conditions (1) -- (9).} but does not have to vanish everywhere on
the $u = 0$ set. Across the set on which $u = 0 \neq R,_r$, the energy density
changes sign. In Sec. \ref{anexample} we will trace all these boundaries on a
numerical example.

\section{There is no geometric singularity at $u/R,_r = 0$}\label{nosing}

\setcounter{equation}{0}

Now we consider the case when $u / R,_r = 0$ at $u = 0$. This is a singularity
of the metric because $\left|g_{11}\right| = \left|{\rm e}^A\right| \to \infty$
there. However, none of the curvature or flow scalars become infinite at those
locations, as we show below.

The components of tensors referred to below are scalar components with respect
to the orthonormal tetrad defined by the metric (\ref{2.1}):
\begin{equation}\label{5.1}
e^0 = {\rm e}^{C/2} {\rm d} t, \qquad e^1 = {\rm e}^{A/2} {\rm d} r, \qquad e^2
= R {\rm d} \vartheta, \qquad e^3 = R \sin \vartheta {\rm d} \varphi.
\end{equation}
These components are thus scalars. In this frame, the nonzero components of the
Ricci tensor and of the scalar curvature are\footnote{We substituted the
solutions of the Einstein -- Maxwell equations for $C$ and $A$ from (\ref{2.5})
-- (\ref{2.9}). These formulae were calculated by the computer-algebra system
Ortocartan \cite{Kras2001}.}
\begin{eqnarray}\label{5.2}
R_{00} &=& \Lambda + \frac G {c^4}\ \left(\frac {Q^2} {R^4} + \frac {u N,_r}
{R^2 R,_r}\right), \nonumber \\
R_{11} &=& R_{22} = R_{33} = - \Lambda + \frac G {c^4}\ \left(- \frac {Q^2}
{R^4} + \frac {u N,_r} {R^2 R,_r}\right), \nonumber \\
R &=& 4 \Lambda - \frac {2 G} {c^4}\ \frac {u N,_r} {R^2 R,_r}.
\end{eqnarray}
The nonzero tetrad components of the Weyl tensor are
\begin{eqnarray}\label{5.3}
C_{0 1 0 1} &=& - C_{2 3 2 3} = - 2W, \nonumber \\
C_{0 2 0 2} &=& C_{0 3 0 3} = - C_{1 2 1 2} = - C_{1 3 1 3} = W, \nonumber \\
W &\df& \frac G {c^4}\ \left(\frac {Q^2} {R^4} + \frac {u N,_r} {3 R^2
R,_r}\right) - \frac {\cal M} {R^3}.
\end{eqnarray}
None of these are singular at $u / R,_r = 0$.

The scalars defined by the flow -- the square of the acceleration vector
$\dot{u}^{\alpha} \dot{u}_{\alpha}$, the expansion $\theta =
{u^{\alpha}};_{\alpha}$ and the shear $\sigma$ given by $\sigma^2 = \frac 1 2
\sigma^{\alpha \beta} \sigma_{\alpha \beta}$, where
\begin{equation}\label{5.4}
\sigma_{\alpha \beta} = u_{(\alpha; \beta)} - \dot{u}_{(\alpha} u_{\beta)} -
\frac 1 3 \theta \left(g_{\alpha \beta} - u_{\alpha} u_{\beta}\right),
\end{equation}
are as follows
\begin{eqnarray}
\dot{u}^{\alpha} \dot{u}_{\alpha} &=& - \frac {Q^2 {Q,_N}^2} {R^4}, \label{5.5}
\\
\theta &=& \frac 1 2 {\rm e}^{- C/2} \left(A,_t + 4 \frac {R,_t} R\right),
\label{5.6} \\
\sigma &=& \frac 1 {2 \sqrt{3}} {\rm e}^{- C/2} \left(2 \frac {R,_t} R -
A,_t\right). \label{5.7}
\end{eqnarray}
The quantities $\dot{u}^{\alpha} \dot{u}_{\alpha}$ and ${\rm e}^{- C/2} R,_t /
R$ are seen to be nonsingular at $u = 0$ (the latter from (\ref{2.7})), so we
investigate ${\rm e}^{- C/2} A,_t$. We find from (\ref{2.5}) and (\ref{2.7}):
\begin{eqnarray}\label{5.8}
{\rm e}^{- C/2} A,_t &=& {\rm e}^{- C/2} \left(\frac {2 R,_{t r}} {R,_r} - \frac
{2 Q Q,_N R,_t} {u R^2}\right) \nonumber \\
&=& \frac {2 {\rm e}^{C/2}} {R,_t R,_r} \left({\rm e}^{- C} R,_t R,_{t r} -
\frac {QQ,_N R,_r} {uR^2}\ {\rm e}^{- C} {R,_t}^2\right) \nonumber \\
&=& \frac {2 {\rm e}^{C/2}} {R,_t} \left\{\left[\Gamma,_r + \frac {GN,_r} {c^4
R} - \frac {{Q,_N}^2 N,_r} R - \frac {QQ,_{NN} N,_r} R\right]\ \frac u
{R,_r}\right. \nonumber \\
&-& \left.\left[\frac {{\cal M} - QQ,_N \Gamma} {R^2} + \frac {Q^2
\left({Q,_N}^2 - G/c^4\right)} {R^3}\right]\right\}.
\end{eqnarray}
This is seen to be nonsingular at $u / R,_r = 0$. Thus, all the flow scalars are
nonsingular at $u / R,_r = 0$, just like all the curvature scalars. This shows
that the singularity in $g_{11} = - {\rm e}^A$ is merely a coordinate
singularity.

Note the following identity
\begin{equation}\label{5.9}
u^2 - \Delta = \left(u^R\right)^2 > 0.
\end{equation}
Thus, at $u = 0$, $\Delta$ must be non-positive. This means, if the region $u
\leq 0$ exists, then the outer surface of this region either lies between the
two R--N event horizons or touches one of them. (As shown before, the boundary
of the region $u \leq 0$ touches the horizons when $u^R = 0$ at $u = 0$.)

It is puzzling that this negative-$u$ region is invisible in the mass-curvature
coordinates. It is not surprising that the coordinate singularity at $u = 0$
disappears -- the mass-curvature coordinates turn thus out to be those that
remove it. However, we showed above that at $u / R,_r = 0$ the mass density
changes sign, and the change of sign of a scalar is an invariant property that
should be visible in all coordinate systems. Equation (\ref{2.20}) shows that
this can happen only by changing the sign of $F$. But then there is no way in
which the $\epsilon$ of (\ref{2.20}) could go through zero -- $F$ is finite
everywhere, except at the turning points, at which $F u^R$ is finite. This seems
to imply that the set $u = 0$ is not covered by the mass-curvature coordinates.

This is indeed likely. Equations (\ref{2.5}) -- (\ref{2.6}) show that
(\ref{2.6}) cannot be integrated across the $u = 0$ set. Then, the definition of
$F$ via the comoving coordinates is $F = R,_r / \left(uu^R {\cal M},_r\right)$,
i.e. the transformation to the $({\cal M}, R)$-coordinates is singular at $u /
R,_r = 0$. Thus, $F$ should suffer a discontinuous jump from positive to
negative values across the set $u / R,_r = 0$, but this change remains invisible
if we use the $({\cal M}, R)$-coordinates throughout.

\section{A vacuole around the center of symmetry is no remedy to $u <
0$}\label{novac}

\setcounter{equation}{0}

Since, with the 9 conditions of Sec. \ref{paper1}, the region $u < 0$ cannot be
eliminated by a choice of the arbitrary functions, can we get rid of it by
matching the charged dust metric to vacuum across a hypersurface $r =$ const
{\it on the inside}? i.e. by cutting an empty vacuole around the center of
symmetry of the charged dust ball? We shall consider the matching only in the
case $\Lambda = 0$.

The electrovacuum spacetime matched to spherically symmetric charged dust must
be Reissner -- Nordstr\"{o}m (R--N). However, if the R--N solution is to be
applied around the center of symmetry, then it will contain a singularity unless
its mass and charge parameters are both zero. In this case, the R--N solution
reduces to the Minkowski spacetime.

Can the charged dust metric be matched to the Minkowski metric, with all the
matching conditions fulfilled? As eq. (\ref{2.11}) implies, the matching
requires that $Q = {\cal M} = 0$ at the matching sphere $r = r_i$. Then,
(\ref{2.7}) shows that if $\Gamma(r_i) = 1$, then $R,_t(t, r_i) = 0$, i.e. the
charged dust particles that are initially on the matching hypersurface will
remain on it permanently. Equation (\ref{2.6}) shows that $C,_r(r_i) = 0$, thus
$C(t, r_i) = C_0(t)$ and can be made zero, so that ${\rm e}^C = 1$, by a
transformation of $t$. This all should happen at a nonzero $R$.

For the dust particles remaining permanently on the matching hypersurface, the
minimal and maximal radius given by (\ref{2.13}) should coincide and both be
nonzero. In order to achieve this, the limit of $- M / E$ at $N = 0$ must be
positive, and in addition, to avoid $R_+$ being zero, the terms $(M/2E)^2$ and
$Q^2 \left({Q,_N}^2 - G / c^4\right)/(2E)$ must be of the same order (so that
their sum at $r = r_i$ does not equal $(M/2E)^2$). These conditions are
impossible to fulfil -- see Appendix \ref{novacu}.

The conditions of a bounce at a nonzero $R$ can be fulfilled in the cases $E
\geq 0$. However, then the arguments used in Refs. \cite{Ori1990, PlKr2006} and
\cite{KrBo2006} still hold: even if $Q = {\cal M} = 0$ occurs at $R > 0$, the
shell crossings will be unavoidable because the function $F$ necessarily changes
sign somewhere in the vicinity of the $Q = 0$ surface.

\section{A transient singularity at the center of symmetry}\label{transient}

\setcounter{equation}{0}

Paper I and the present paper were concerned with avoiding the BB/BC, shell
crossings and permanent central singularities. It turns out that there is one
more kind of singularity that has slipped through all the tests applied so far.

We noted in Sec. \ref{negucons} that $R \to 0$ faster than anywhere else if we
approach the center of symmetry along the hypersurface $S_{\rm min}$, in which
$R = R_+$. The energy density given by (\ref{2.10}) will be finite at the center
provided the product $u N,_r / \left(R^2 R,_r\right) < \infty$. We have already
shown in Sec. \ref{negdens} that $u$ is finite at all points of the world line
of the center, albeit negative when the center is approached from within $S_{\rm
min}$. Thus, to avoid a singularity at all points of the center, we must have
$\lim_{r \to r_c} N,_r / \left(R^2 R,_r\right) \equiv \lim_{r \to r_c} 3 /
\left(R^3\right),_N < \infty$, i.e. $\lim_{r \to r_c} \left(R^3\right),_N \neq
0$. However, as we show below, $\lim_{r \to r_c} \left(R^3\right),_N = 0$ if the
center of symmetry is approached along the hypersurface $S_{\rm min}$, and also
$\lim_{r \to r_c} \left(R^3\right),_N = 0$ if the hypersurface $S_{\rm min}$ is
approached along the center of symmetry.

A nonzero limit of $\left(R^3\right),_N$ at the center, where $R = 0 = N$, means
that $R^3$ must tend to zero as fast as $N$, i.e. that $R$ must tend to zero as
fast as $N^{1/3}$ and no faster. Let us then take the first of (\ref{2.27}). The
sufficient condition for $\lim_{r \to r_c} R/N^{1/3} \neq 0$ is then
\begin{equation}\label{7.1}
\lim_{N \to 0} \frac M {2E N^{1/3}} \neq 0,
\end{equation}
which does the job at those points where $\cos \omega \neq 1$. With $\cos \omega
= 1$, however, where $R = R_+$, we have
\begin{equation}\label{7.2}
R = R_+ = \frac M {(- 2E)} \left[1 - \sqrt{1 - \frac {2EQ^2 \left({Q,_N}^2 - G /
c^4\right)} {M^2}}\right].
\end{equation}
We proved in Appendix \ref{negli} that the fraction under the square root always
tends to zero at $r \to r_c$ if the other regularity conditions are obeyed.
Thus, the expression in square brackets also tends to zero at $r \to r_c$,
implying that $R / N^{1/3} \llim{r \to r_c} 0$, i.e. a singularity at the
center.

We can approach the singular point along the world line of the center of
symmetry. Assuming that $\lim_{r \to r_c} M / \left(- 2E N^{1/3}\right) =
\beta_0 \neq 0$, we obtain
\begin{equation}\label{7.3}
\lim_{r \to r_c} \frac {R(t, r)} {N^{1/3}} = \beta_0 (1 - \cos \omega) \neq 0
\end{equation}
(the dependence on time is now hidden in $\omega$). Taking the limit of
(\ref{7.3}) as $\cos \omega \to 1$, we again obtain the zero limit of $R /
N^{1/3}$ at the feral point.

Let us recall: this transient singularity results because the $R_+$ given by
(\ref{7.2}) goes to zero at $r \to r_c$ faster than $N^{1/3}$. Can we avoid the
singularity by forcing $R_+$ to go to zero at $r \to r_c$ as fast as $N^{1/3}$
or slower? The following simple consideration provides a negative answer.
Equation (\ref{7.2}) can be equivalently written as
\begin{equation}\label{7.4}
R_+ = - \frac {Q^2 \left({Q,_N}^2 - G / c^4\right)} {M \left[1 + \sqrt{1 -
{\displaystyle \frac {2EQ^2 \left({Q,_N}^2 - G / c^4\right)} {M^2}}}\right]}.
\end{equation}
We can now use the reasoning in Appendix \ref{negli} to show that $\lim_{N \to
0} R_+/N^{1/3} = 0$ independently of the forms of the functions $Q$, $E$ and
$M$, provided conditions (1) -- (9) of sec. 2 are obeyed. Equation (\ref{b.3})
tells us what forms $p$ and $\left({p,_x}^2 - 1\right)$ must have in a vicinity
of the center. Thus we have
\begin{eqnarray}\label{7.5}
&-&\frac {Q^2 \left({Q,_N}^2 - G / c^4\right)} M \nonumber \\
&=& \frac {x^2 \left(1 + A_p x^{\alpha_p - 1} + {\cal O}_{\alpha_p - 1}\right)^2
\left(2 A_p x^{\alpha_p - 1} + {\cal O}_{\alpha_p - 1}\right)} {A_p
\left(\alpha_p + 1\right) x^{\alpha_p} + {\cal O}_{\alpha_p}} \nonumber \\
&=& \frac {2x} {\alpha_p + 1} + {\cal O}_1 \equiv \frac {2N} {N_0 \left(\alpha_p
+ 1\right)} + {\cal O}_1.
\end{eqnarray}
Because of (\ref{7.5}) and (\ref{7.1}), the expression under the square root in
(\ref{7.4}) has the limit 1 at $r \to r_c$. Consequently
\begin{equation}\label{7.6}
\frac {R_+} {N^{1/3}} = \frac {N^{2/3}} {N_0 \left(\alpha_p + 1\right)} + {\cal
O}_{2/3} \llim{N \to 0} 0,
\end{equation}
$\square$.

Note that since $u < 0$ as the singularity is approached from within $S_{\rm
min}$, and at the center necessarily $R > 0$ and $R,_N > 0$, the density in the
singularity becomes {\it minus infinity}. At the same time, if we start at a
point of the center of symmetry different from the maximal compression instant,
then, from (\ref{2.21}) -- (\ref{2.24}) and ${Q,_N}^2(r_c) = G / c^4$ we have
$u(r_c) = 1 > 0$ everywhere on $N = 0$, so $\epsilon > 0$ everywhere on $N = 0$,
including the limiting point of maximal compression. Consequently, $\epsilon \to
+ \infty$ at the point where the center of symmetry hits the maximal compression
hypersurface. This singularity is thus direction-dependent.

This transient singularity exists within each $S_{\rm min}$ hypersurface.
However, with all other conditions of regularity obeyed, the limit of the period
of oscillations of the solution (\ref{2.27}) at $r \to r_c$ is infinite, as will
be shown below. Thus, the transient singular point can be reached only once,
where the bounce set $s = s_B({\cal M})$ hits the center of symmetry at a finite
$s$. Every other hypersurface $S_{\rm min}$ escapes to the infinite future or
the infinite past when it approaches the center of symmetry.

Here is the proof that $T_p \llim{r \to r_c} \infty$, where $T_p$ is the period
of oscillation given by (\ref{2.28}). We know that $M / (- 2E)$ must tend to
zero at the center as fast as $N^{1/3}$. We know from (\ref{2.33}) that $M$
tends to zero faster than $N$, so let $M \propto N^{1 + \varepsilon}$, where
$\varepsilon > 0$. Consequently, $(- 2E)$ tends to zero as fast as $N^{2/3 +
\varepsilon}$. This means that $T_p = M / (- 2E)^{3/2}$ behaves as $N^{-
\varepsilon / 2}$, i.e. tends to infinity at the center. $\square$

Since we were interested in sending a ball of dust through the Reissner --
Nordstr\"{o}m throat, we set up the initial conditions so that the state of
maximal compression (minimal size) was attained simultaneously by all mass
shells. In such a configuration, illustrated in Figs. 12 and 13 of Paper I (and
also in Fig. \ref{drawcycles2} further here), the hypersurfaces $S_{\rm max}$
(consisting of the instants of maximal size of all the shells) approach the
center of symmetry in the infinite past and in the infinite future.
Consequently, it is not in general guaranteed that the center of symmetry will
ever reach the region where $u > 0$ and $\epsilon > 0$, but in our explicit
example, given in the next section, the region $u < 0$ is contained in a small
neighbourhood of $S_{\rm min}$.

However, we could take exactly the same example and set up the initial
conditions so that the state of minimal compression (maximal size) is attained
simultaneously. Then the dust ball would never emerge from the R--N throat
because shell crossings would appear immediately after the bounce. However,
initially and for some (perhaps infinite) time to the future and to the past,
the center of symmetry would be in the region of positive $u$ and $\epsilon$.

\section{An example}\label{anexample}

\setcounter{equation}{0}

To illustrate the statements of the preceding sections we will use the same
example that we introduced in Paper I. To recall, in the example the function
$N$ was (and still will be) used as the radial coordinate. In the exemplary
configuration, the charge function was
\begin{equation}\label{8.1}
Q(N) = q \frac {\sqrt{G} N_0} {c^2}\ p(x),
\end{equation}
where $q = \pm 1$, to allow for any sign of the charge, $x \df N/N_0$, $N_0$ is
an arbitrary constant, and
\begin{equation}\label{8.2}
p(x) =  x / (1 + x)^2.
\end{equation}
Then
\begin{equation}\label{8.3}
Q,_N = q \frac {\sqrt{G}} {c^2}\ \frac {1 - x} {(1 + x)^3},
\end{equation}
and
\begin{equation}\label{8.4}
F_1(x) \df 1 - \frac {c^4} G\ \left({Q,_N}^2 + QQ,_{NN}\right) = 1 - \frac {3x^2
- 6x + 1} {(1 + x)^6}.
\end{equation}

The original definition of the function $E(x)$ was incorrect, as it implied a
permanent central singularity. We quote here the corrected $E(x)$ and the
formulae dependent on it, from the gr-qc version of Paper I. The functions
defined along the way obey conditions (1) -- (9) of Sec. \ref{paper1} -- for
proofs see the gr-qc version of Paper I. Thus
\begin{equation}\label{8.5}
2E = - \frac {b x^{5/3}} {1 + b x^{5/3}},
\end{equation}
where $b$ is another arbitrary constant. With such $E$ we have
\begin{equation}\label{8.6}
\Gamma(x) = \frac 1 {\sqrt{1 + b x^{5/3}}},
\end{equation}
\begin{equation}\label{8.7}
{\cal M}(x) = \frac {GN_0} {c^4} \int_0^x \frac {{\rm d} x'} {\sqrt{1 + b
{x'}^{5/3}}} \df \frac {GN_0} {c^4} \mu(x).
\end{equation}
Then, further:
\begin{eqnarray}\label{8.8}
M &\equiv& {\cal M} - QQ,_N \Gamma = \frac {GN_0} {c^4}\ F_2(x), \nonumber \\
F_2(x) &\df& \mu(x) - \frac {x (1 - x)} {(1 + x)^5 \sqrt{1 + b x^{5/3}}}.
\end{eqnarray}
For checking condition (7) we need the function $F_3(x)$ defined below. This
condition is equivalent to
\begin{equation}\label{8.9}
F_3(x) > 0, \qquad F_3(x) \df {F_2}^2(x) - F_6(x),
\end{equation}
where
\begin{equation}\label{8.10}
F_6(x) \df - 2E p^2 \left(1 - {p,_x}^2\right) = \frac {b x^{11/3}} {\left(1 + b
x^{5/3}\right) (1 + x)^4}\ \left[1 - \frac {(1 - x)^2} {(1 + x)^6}\right].
\end{equation}
Fig. \ref{chadu6fig} shows the graphs of the functions defined above with $b =
2.5$ (why this value -- see below). This is a corrected version of Fig. 7 from
Paper I.

 \begin{figure}
 \begin{center}
 \includegraphics[scale = 0.9]{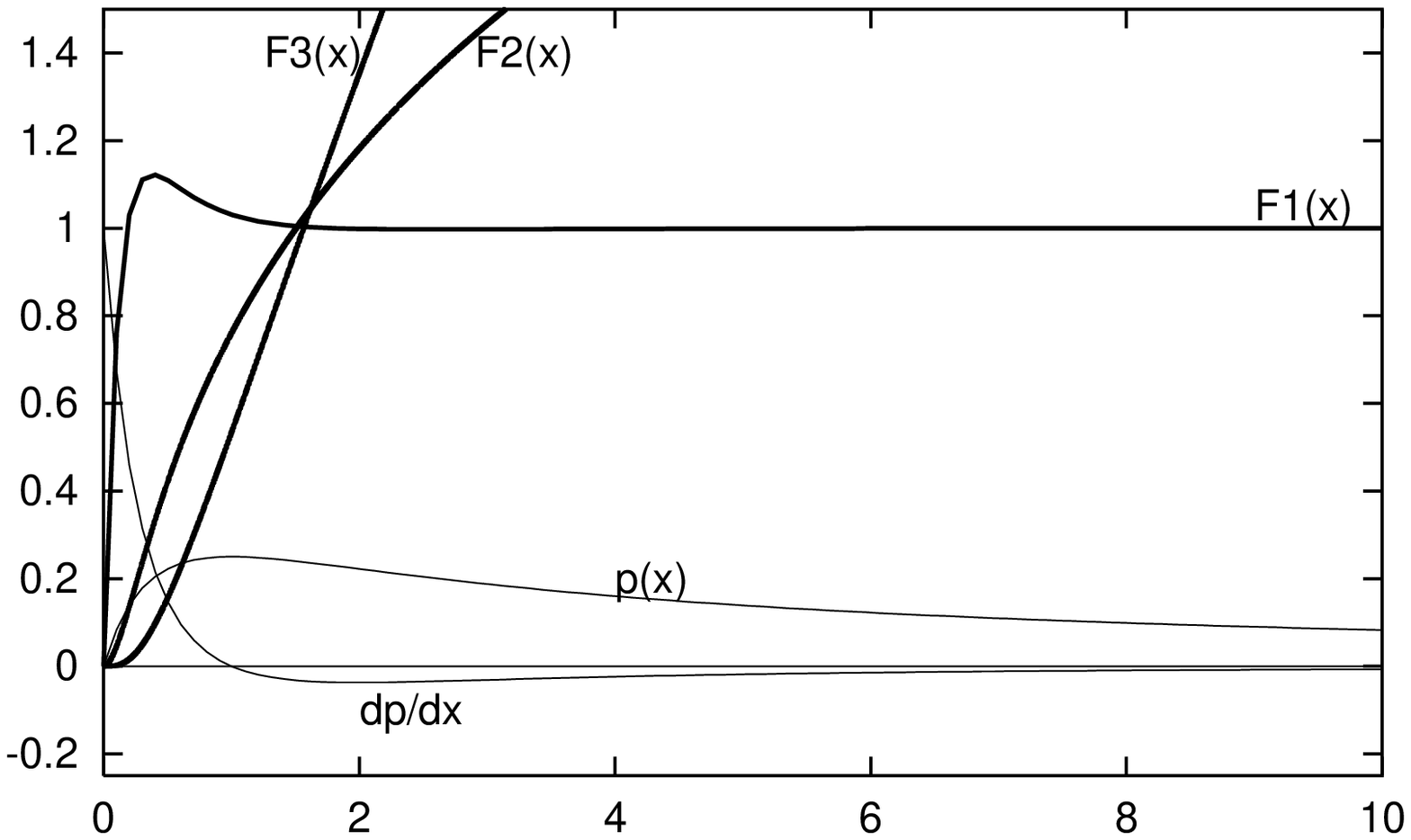}
 ${}$ \\[-45mm]
 \hspace{70mm}
  \includegraphics[scale = 0.35]{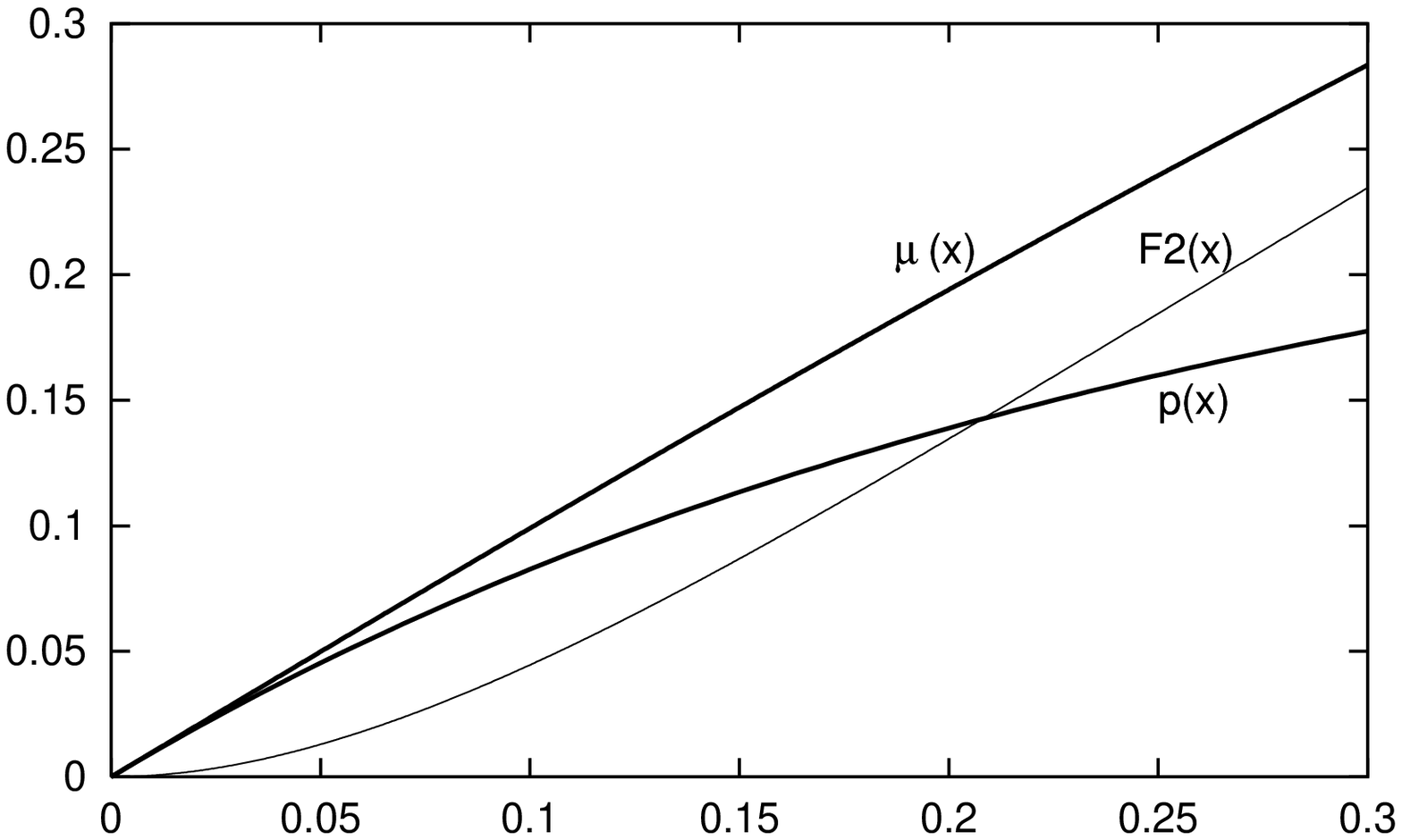}
 ${}$ \\[20mm]
 \caption{
 \label{chadu6fig}
 \footnotesize
Graphs of the functions $p(x)$, $\dril p x$, $F_1(x)$, $F_2(x)$ and $F_3(x)$.
The inset shows the functions $p(x)$, $F_2(x)$ and $\mu(x)$ in a vicinity of $x
= 0$.}
 \end{center}
 \end{figure}

\begin{figure}
 \begin{center}
 \includegraphics[scale = 0.8]{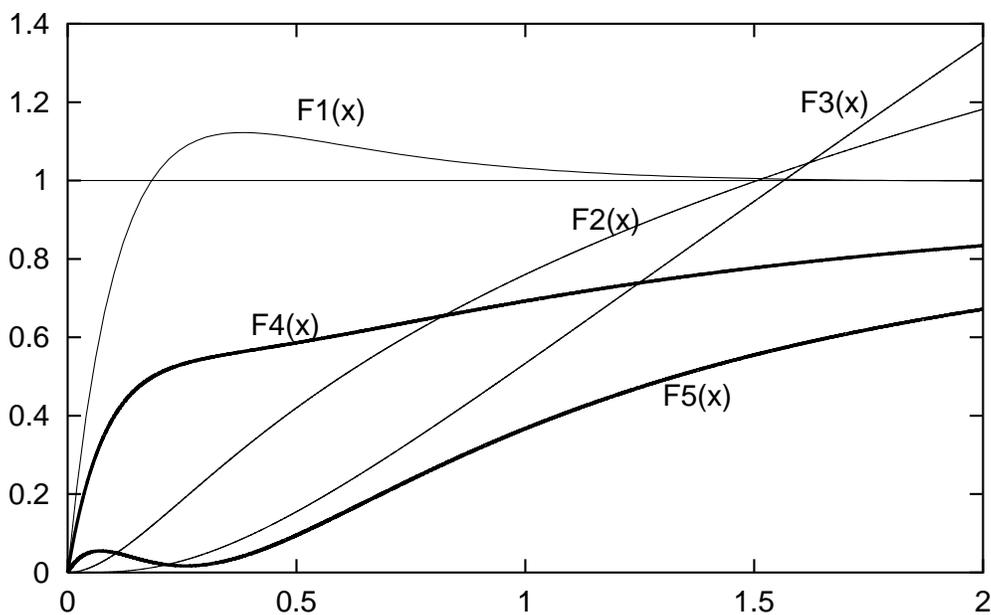}
 \caption{
 \label{chadu7fig}
 \footnotesize
Graphs of the functions $F_4(x)$ and $F_5(x)$ with $b = 2.5$. The graphs of
$F_1(x)$, $F_2(x)$ and $F_3(x)$ are shown for comparison and for scale.}
 \end{center}
 \end{figure}

For checking the remaining conditions we need the functions $F_4(x)$ and
$F_5(x)$ defined below. Condition (8) is equivalent to:
\begin{equation}\label{8.11}
F_4(x) > 0, \qquad {\rm where} \qquad F_4(x) \df F_1(x) - \frac 5 {6x \sqrt{1 +
b x^{5/3}}}\ F_2(x),
\end{equation}
and condition (9) is equivalent to:
\begin{equation}\label{8.12}
F_5(x) > 0, \qquad {\rm where}\ F_5(x) \df F_1(x) - \frac 5 {6 x \sqrt{1 + b
x^{5/3}}} \left[F_2(x) + \sqrt{F_3(x)}\right].
\end{equation}
Fig. \ref{chadu7fig} shows the graphs of $F_4(x)$ and $F_5(x)$ with $b = 2.5$.

The reason for choosing $b = 2.5$ was this: the graph of $F_5$ is sensitive to
the value of $b$. With decreasing $b$, the local minimum of $F_5$ becomes
smaller, and for $b$ small enough (for example, $b = 0.75$) $F_5 < 0$ around the
minimum. With $b \geq 2.5$, the minimum is clearly positive. The value $b = 2.5$
was used in all the subsequent figures.

\begin{figure}
 \begin{center}
 \includegraphics[scale = 0.8]{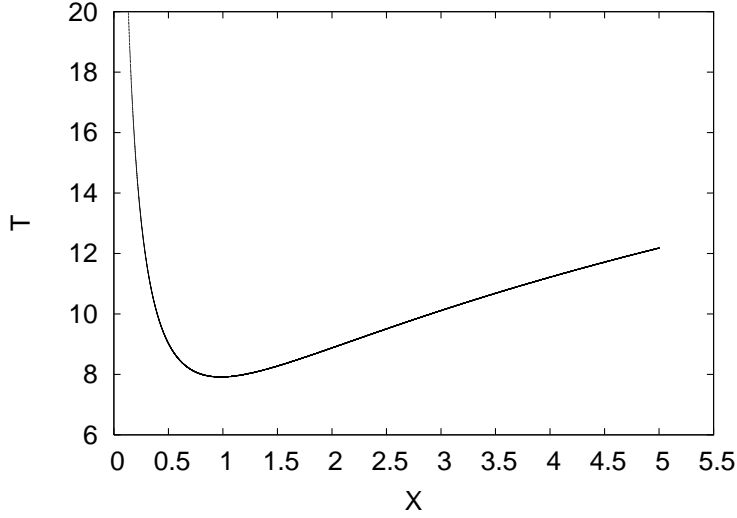}
 \caption{
 \label{plotperiod}
 \footnotesize
The period (in the time coordinate) as a function of $x = N/N_0$.}
 \end{center}
 \end{figure}

\begin{figure}
 \begin{center}
 \includegraphics[scale = 0.8]{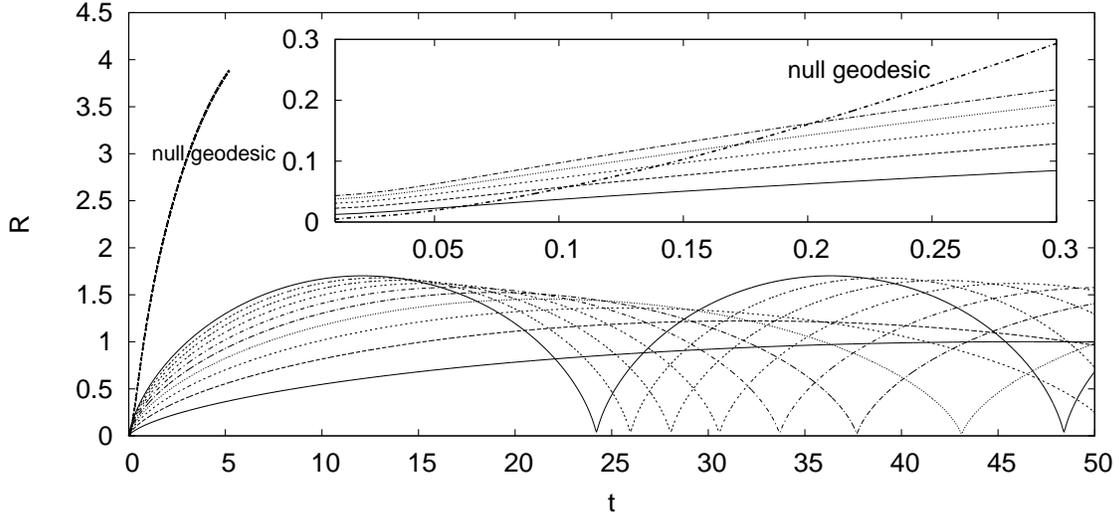}
 \caption{
 \label{drawcycles2}
 \footnotesize
The curves $R({\cal M}, t)$ corresponding to several values of ${\cal M}$. The
mass increases uniformly from $x = 0.01$ on the lowest curve to $x = 0.1$ on the
highest curve. The origin $t = R = 0$ is the point singularity discussed in Sec.
\ref{transient}. A null cone with the vertex at the point singularity is also
shown -- as can be seen, by the time the outermost mass shell reaches its
maximal size, it is within the future light cone of the singularity. The inset
shows the light cone in a small neighbourhood of its vertex -- all mass shells
enter this cone soon after going through the bounce. The region $u < 0$, as
follows by comparison with Fig. \ref{u=0}, would occupy a very small
neighbourhood of the origin and would not be visible at the scale of this
figure. }
 \end{center}
 \end{figure}

Figs. 9 and 10 of Paper I do not change in any noticeable way after the
correction done in $E(x)$, so they need not be repeated here (see the gr-qc
version of Paper I). However, Figs. 11 and 12 do change significantly (also in
consequence of the different value of $b$ adopted here), and their correct
versions are shown in Figs. \ref{plotperiod} and \ref{drawcycles2} here. Fig.
\ref{drawcycles2} also shows the light cone of the transient singularity at $(t,
R) = (t_B, 0)$ -- to demonstrate that all mass shells enter this cone soon after
going through the minimal size.

\begin{figure}
 \begin{center}
 \hspace{-90mm}
 \includegraphics[scale = 0.6]{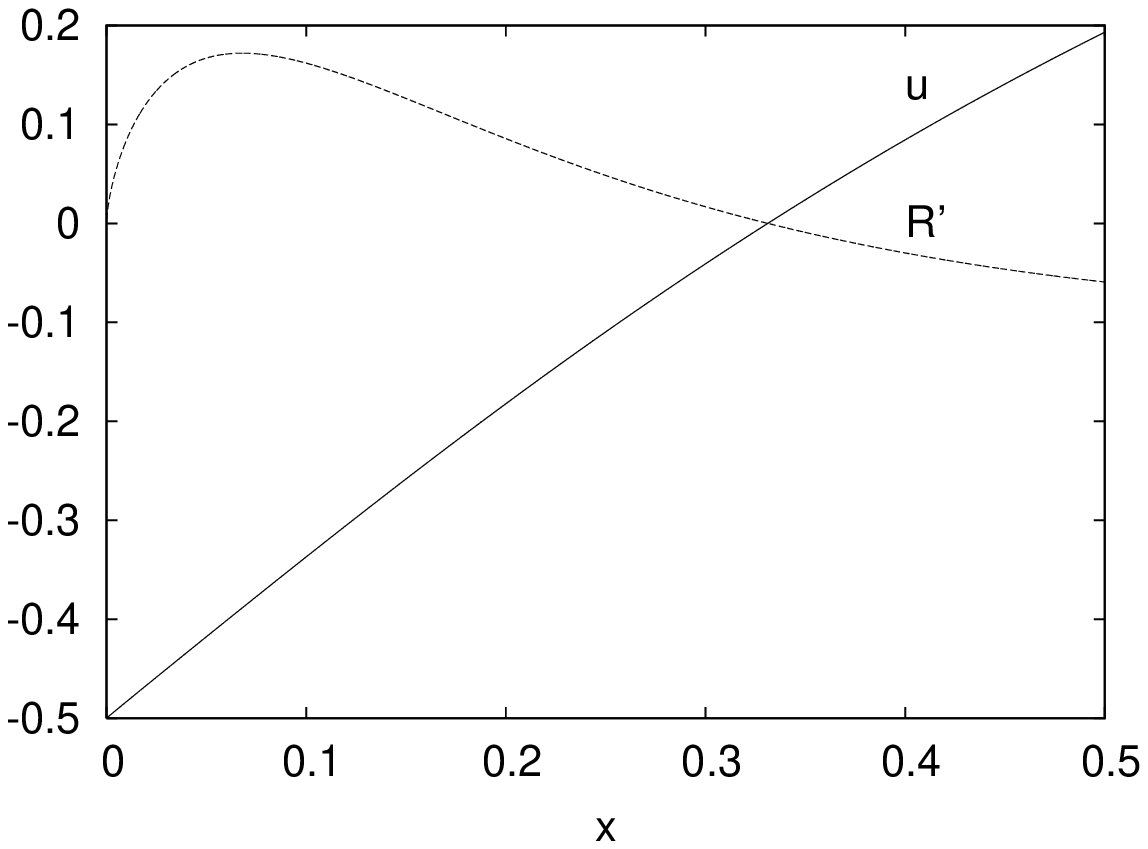}
 ${}$ \\[-54mm]
 \hspace{70mm}
  \includegraphics[scale = 0.62]{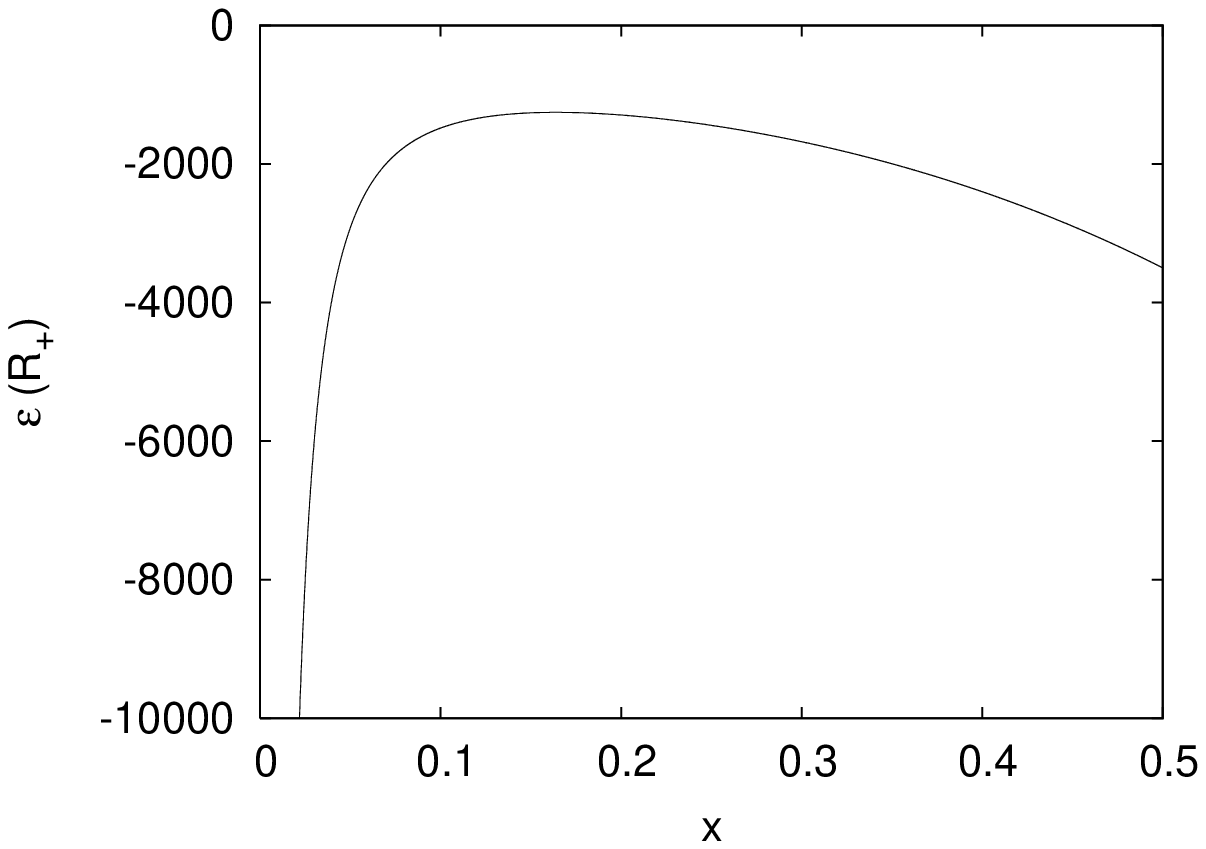}
 \caption{
 \label{uRpratR+}
 \footnotesize
The  functions $u(x)$ and $R,_r(x)$ (left graph) and energy density
$\epsilon(x)$ (right graph) in the surface $R = R_+$. The first two change their
signs at the same value of $x$, which causes that $u / R,_r < 0$ in this whole
surface, implying $\epsilon < 0$ in the whole surface. The inset in the left
graph shows the neigbourhood of the intersection point. Note that $R,_r \to 0$
at $r \to 0$, which causes the singularity at the center. As shown in Secs.
\ref{negucons} and \ref{transient}, the energy density in this surface is
negative in a vicinity of the center and goes to $- \infty$ at the center.}
 \end{center}
 \end{figure}

The other qualitative conclusions presented in Paper I remain unchanged. Below,
we present the new results found in the present paper. In all the subsequent
figures, the radial coordinate $r$ was defined as $r = N^{1/3}$, so that one can
instantly see whether there is a central singularity ($R,_r = 0$ at the center)
or not ($R,_r > 0$ at the center). Fig. \ref{uRpratR+} shows the functions
$u(x)$ and $R,_r(x)$ (left graph) and the energy density $\epsilon(x)$ (right
graph) in the surface $R = R_+$. The functions $u(x)$ and $R,_r(x)$ have a zero
at the same value of $x$ and opposite signs in all other points, except at $x =
0$ where $R,_r \to 0$ and $u < 0$, so $u / R,_r \to - \infty$. As follows from
the calculations in Secs. \ref{negucons} and \ref{transient}, in this surface
the energy density is negative in a vicinity of the center and goes to $-
\infty$ at the center.

The fact that the point where $u = 0 = R,_r$ does not seem to be singular is
rather mysterious because it is a limiting point of the contour $R,_r = 0$ in
the $(t, x)$-plane, and the rest of the contour is a shell crossing (see Fig.
\ref{u=0}), where the energy density does go to $\pm \infty$. But the curves in
Fig. \ref{drawcycles2} are all for smaller values of $x$, so the singularity is
not in its range.

\begin{figure}
 \begin{center}
 \hspace{-90mm}
 \includegraphics[scale = 0.6]{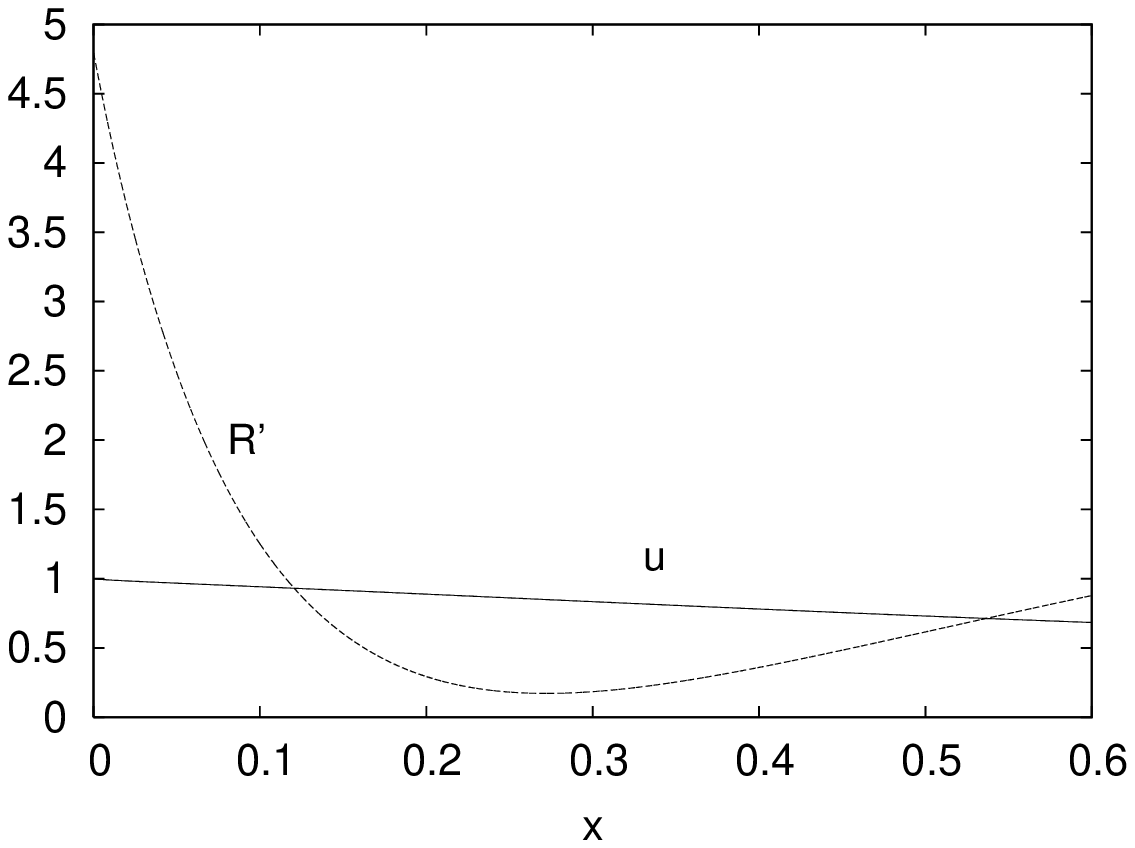}
 ${}$ \\[-54mm]
 \hspace{70mm}
  \includegraphics[scale = 0.62]{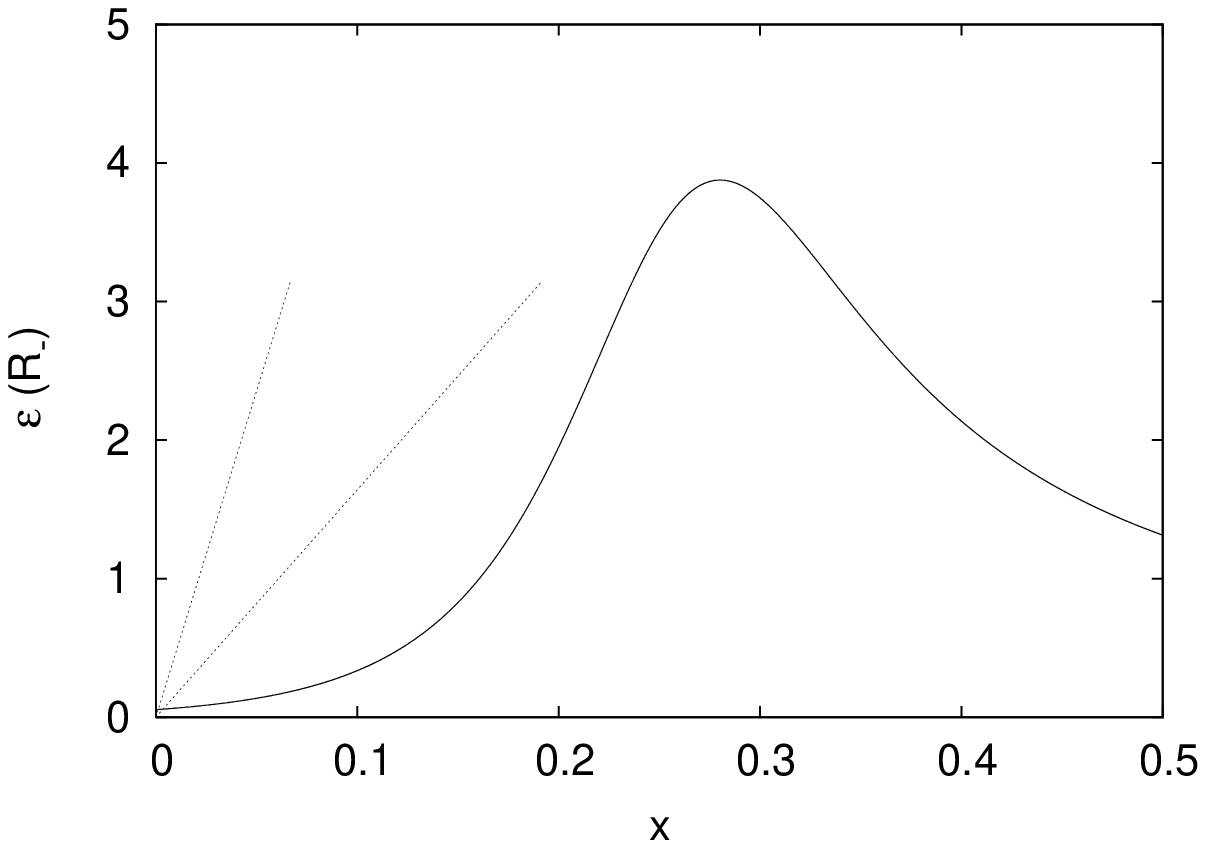}
 \caption{
 \label{uRpratR-}
 \footnotesize
The functions $u(x)$ and $R,_r(x)$ (left graph) and energy density $\epsilon(x)$
(right graph) in the surface $R = R_-$. The functions $u(x)$ and $R,_r(x)$ are
positive everywhere, implying $\epsilon > 0$ in the whole surface. This time
$R,_r > 0$ at $r = 0$, i.e. there is no singularity at the center -- as
confirmed in the right graph. The inset in the right graph shows that the energy
density at the center is not zero, but has a small positive value. }
 \end{center}
 \end{figure}

\begin{figure}
 \begin{center}
 \includegraphics[scale = 0.8]{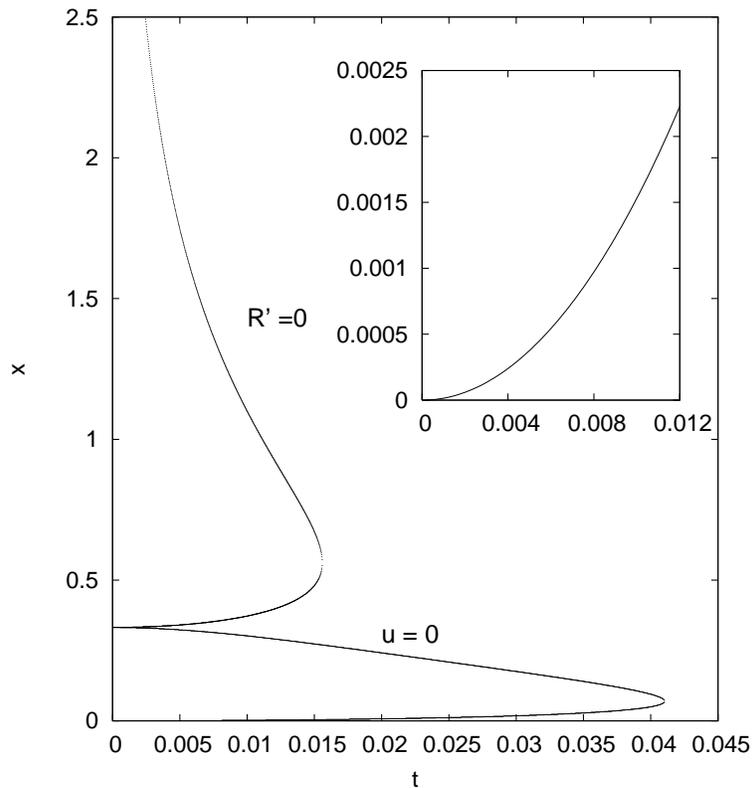}
 \caption{
 \label{u=0}
 \footnotesize
The $u = 0$ and $R,_r = 0$ curves in the $(t, x)$ plane. Each of the functions
$u$ and $R,_r$ is negative to the left of its graph and positive everywhere to
the right. The full contour is mirror-symmetric with respect to the $x$-axis.
The $R,_r = 0$ curve is the locus of shell crossings. The inset shows a closeup
view of the neighbourhood of the origin -- to demonstrate that the $u = 0$ curve
hits the $t$-axis only at $t = 0$. Thus, at the center of symmetry, $u > 0$
everywhere except at the singular origin (where its value becomes dependent on
the path of approach). See more explanation in the text. }
 \end{center}
 \end{figure}

Fig. \ref{uRpratR-} shows the functions $u(x)$ and $R,_r(x)$ (left graph) and
the energy density $\epsilon(x)$ (right graph) in the surface $R = R_-$. Within
this surface $u(x)$ and $R,_r(x)$ are everywhere positive.\footnote{We recall
that the surface $R = R_-$ in our example never intersects the worldline of the
center of symmetry, but approaches it asymptotically as $t \to + \infty$ or $t
\to - \infty$, depending on whether it lies to the future or to the past of $R =
R_+$.} The energy density is everywhere positive and finite (it does not tend to
zero, but to a small positive value as $r \to 0$, as shown in the inset).

Fig. \ref{u=0} shows the curves $u = 0$ (lower part) and $R,_r = 0$ (upper part)
in the $(t, x)$-plane (actually, each curve is one half of the full contour,
which is mirror-symmetric with respect to the $x$-axis). The function $u$ is
negative to the left of the lower contour and positive to the right of it. The
function $R,_r$ is negative to the left of the upper contour and positive to the
right of it. As stated above, the set $R,_r = 0$ is a shell crossing. Its
presence proves that the 9 necessary conditions listed at the end of Sec.
\ref{paper1} were not sufficient. Note the characteristic features of the
curves, consistent with the calculations of Secs. \ref{negucons} and
\ref{nosing}:

1. The lower curve hits the $t$-axis (the center of symmetry) at $t = 0$ (in the
figure, we chose $t_B = 0$, so $t = 0$ is the simultaneously achieved state of
minimal size). Thus $u$ remains positive all the time as we proceed toward $t =
0$ along the center of symmetry.

2. This curve intersects the $x$-axis at some $x > 0$. This intersection point
coincides with the point where the $u(x)$ and $R,_r(x)$ curves intersect in Fig.
\ref{uRpratR+}. The upper curve begins at the same point.

3. To the right of the lower curve, we have $u > 0$, and to the right of the
upper curve $R,_r > 0$. The energy density is positive in the area to the right
of both curves, and negative to the left of any of them, including the $x$-axis.

\begin{figure}
 \begin{center}
 \includegraphics[scale = 0.63]{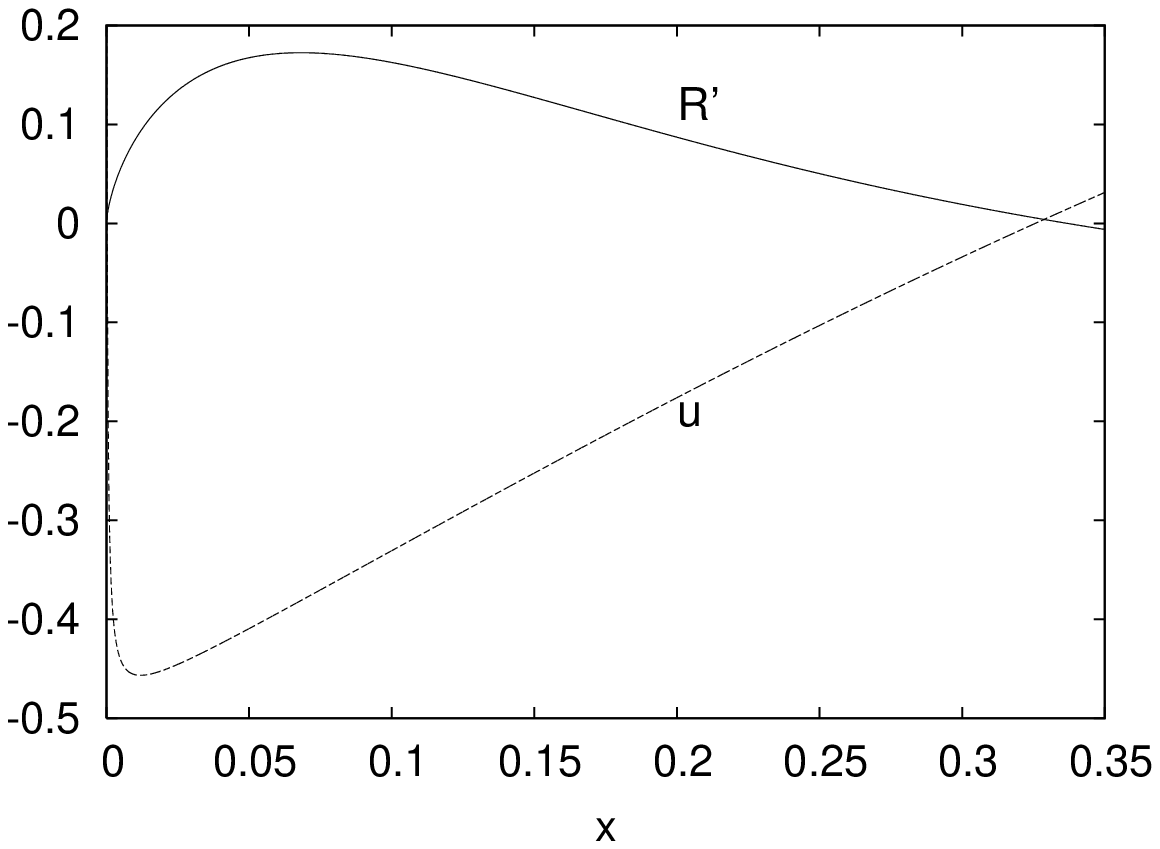}
  \includegraphics[scale = 0.63]{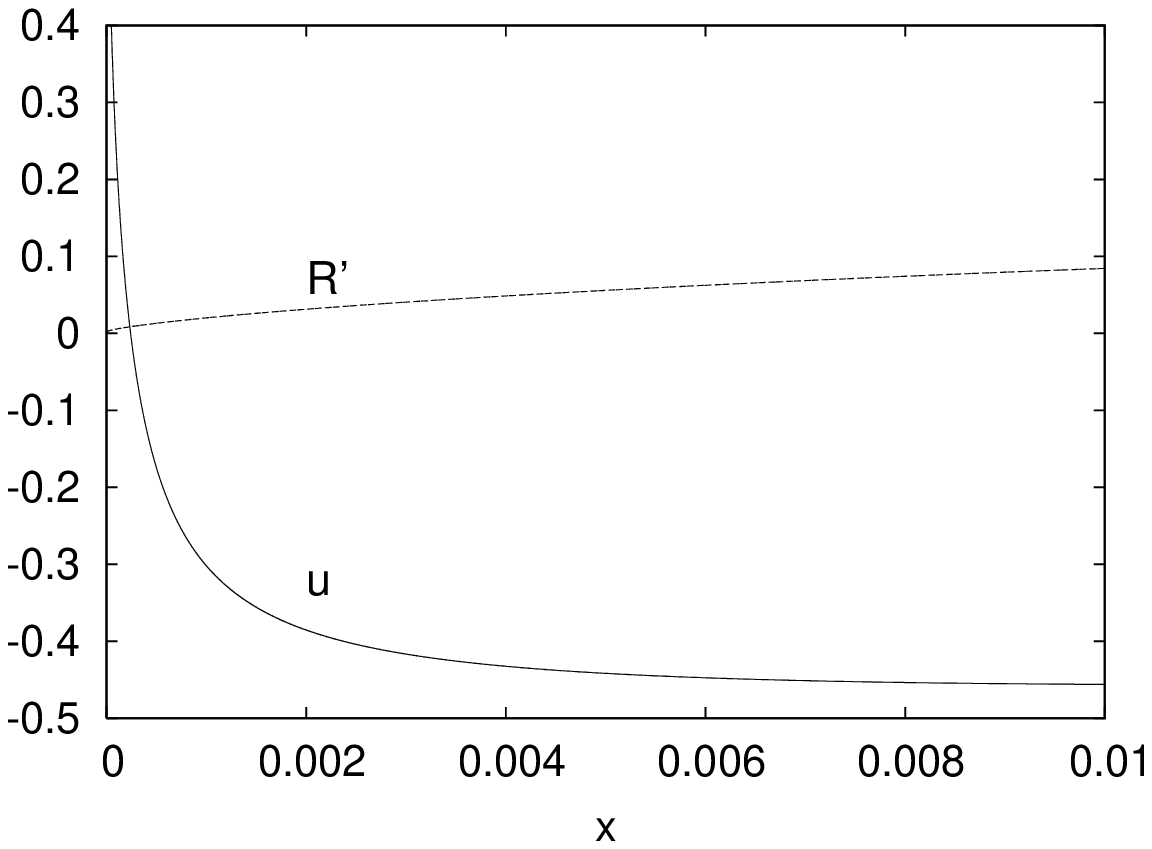}
 \caption{
 \label{uRpratsmalt}
 \footnotesize
The functions $u(x)$ and $R,_r(x)$  along the line $t = 0.004$ in Fig.
\ref{u=0}. The left graph presents these functions for larger values of x, the
graph on the right for smaller values of x. Note that $u$ increases to a
positive value at the center $x = 0$, even though it is negative in a vicinity
of the center.}
 \end{center}
 \end{figure}

\begin{figure}
 \begin{center}
\hspace{-3mm}
  \includegraphics[scale = 0.55]{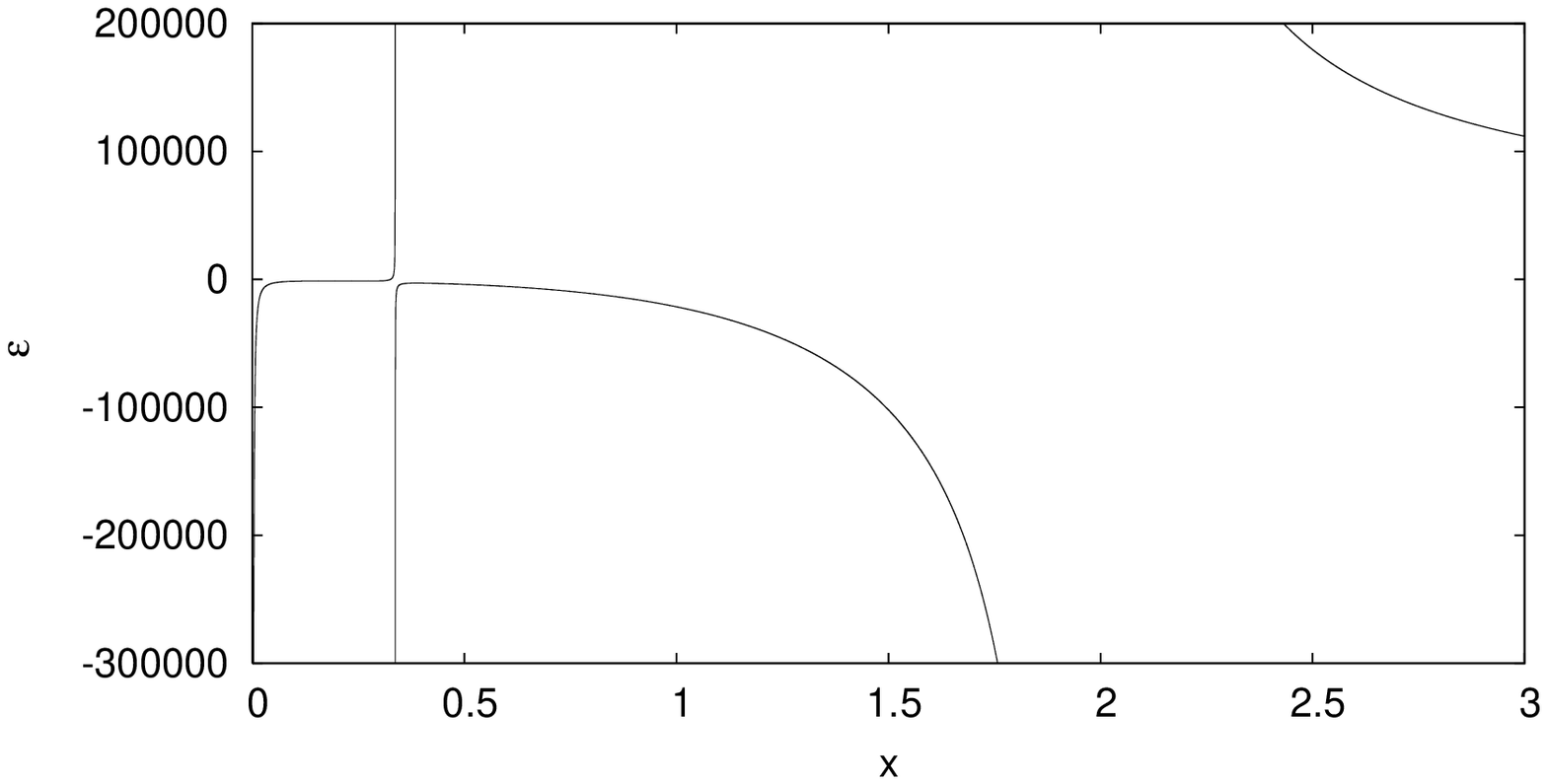}
\hspace{-7mm}
  \includegraphics[scale = 0.55]{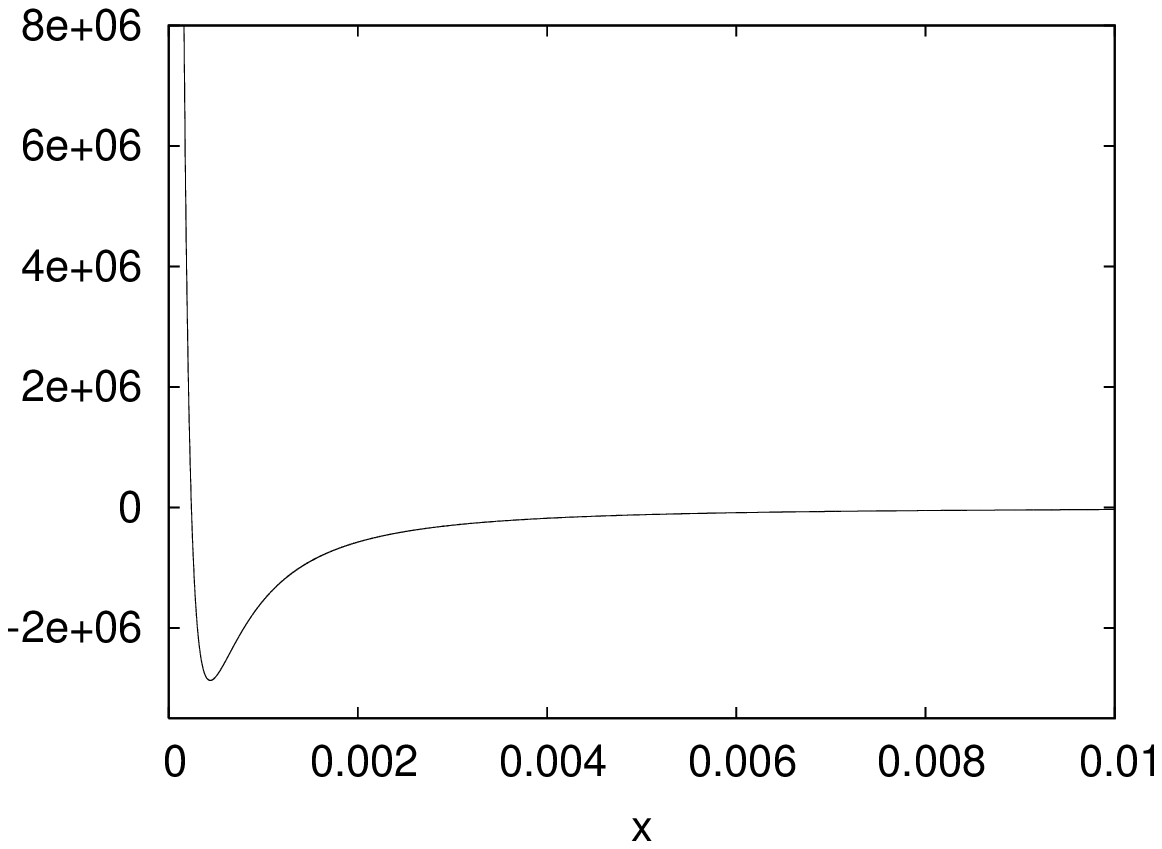}
\caption{
 \label{den2atsmalt}
 \footnotesize
The energy density $\epsilon(x)$ along the line $t = 0.004$ in Fig. \ref{u=0}.
The left graph presents $\epsilon(x)$ for large values of x, the graph on the
right for small values of x. Note, in the left graph, that the curve goes
through zero as it crosses the $u = 0$ contour, but very near to that point it
crosses the $R,_r = 0$ contour which is a shell crossing. There, it goes to
$+\infty$ on one side and to $- \infty$ on the other. The second intersection
with $R,_r = 0$ is also seen at $x \approx 2$, where the density changes sign
again by a jump from $-\infty$ to $+ \infty$. The right graph shows how the
density becomes negative with very large absolute value in a vicinity of the
center at $x = 0$, but then, still closer to the center, grows to very large
positive values.}
 \end{center}
 \end{figure}

In order to better visualise the variation of the functions $u$, $R,_r$ and
$\epsilon$ in the $(t, R)$-plane, we provide three further graphs. Fig.
\ref{uRpratsmalt} shows the functions $u(x)$ and $R,_r(x)$ along the line $t =
0.004$ in Fig. \ref{u=0}. Fig. \ref{den2atsmalt} shows the energy density along
the same section. It changes sign every time when the line $t = 0.004$ crosses
one contour or the other. Where $u = 0$, it changes from negative to positive by
smoothly going through zero; where $R,_r = 0$ it changes sign by jumping from $+
\infty$ to $- \infty$ or the other way round.

\begin{figure}
 \begin{center}
 \hspace{-80mm}
 \includegraphics[scale = 0.6]{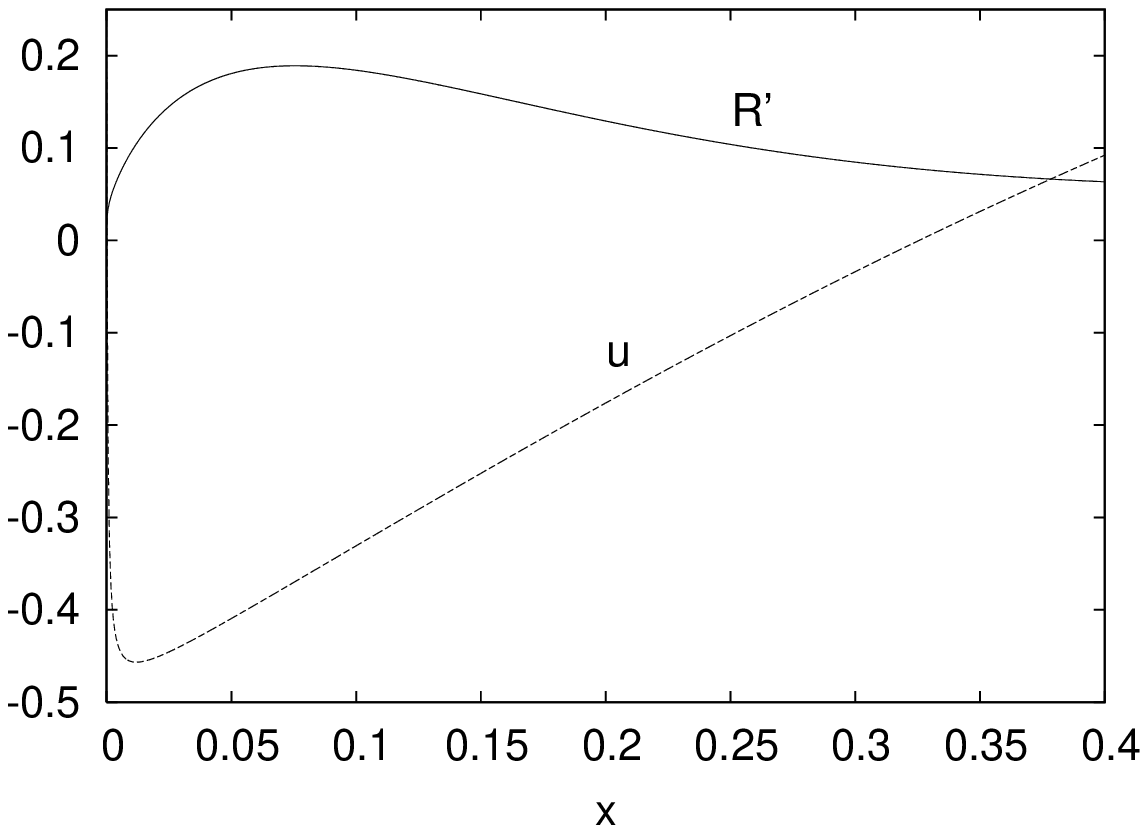}
 ${}$ \\[-54mm]
 \hspace{60mm}
 \includegraphics[scale = 0.62]{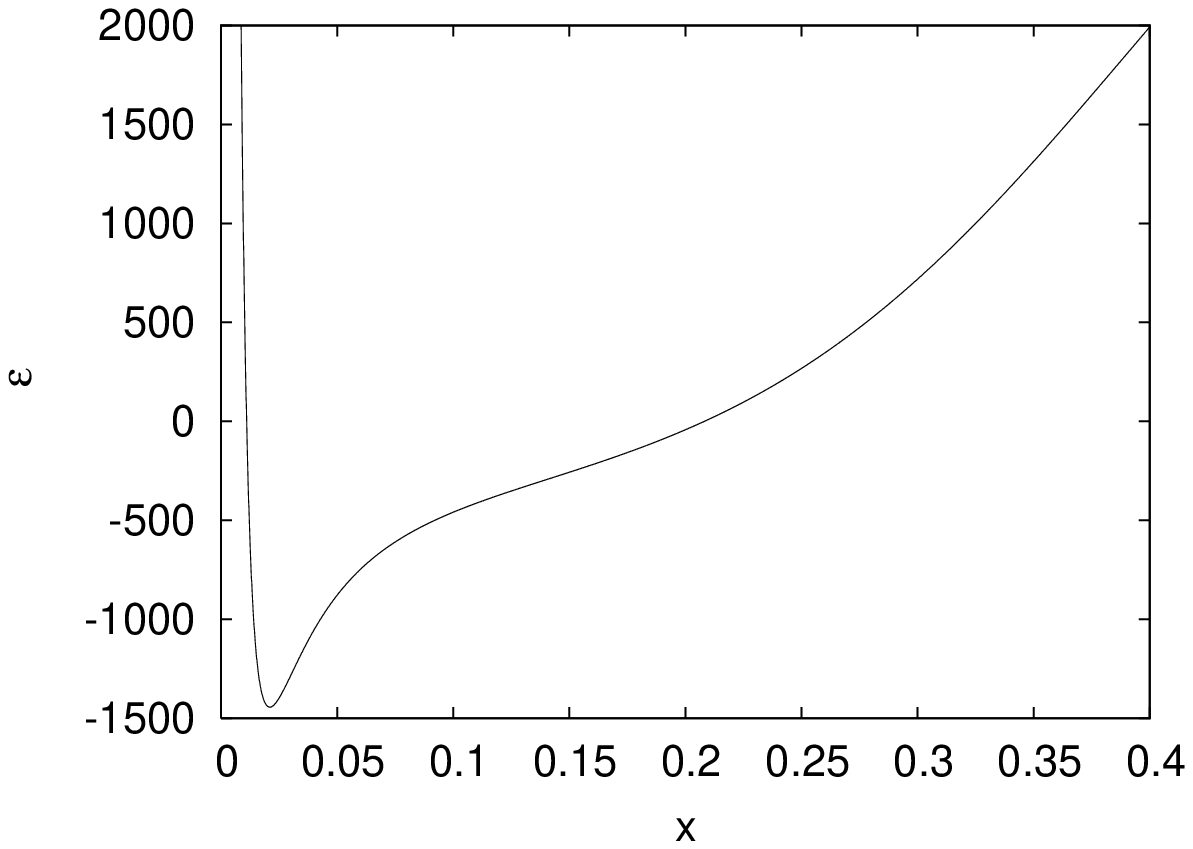}
 \caption{
 \label{uRpratmidt}
 \footnotesize
The functions $u(x)$ and $R,_r(x)$ (left graph) and the energy density
$\epsilon(x)$ (right graph)  along the line $t = 0.025$ in Fig. \ref{u=0}. }
 \end{center}
 \end{figure}

\begin{figure}
 \begin{center}
 \includegraphics[scale = 0.8]{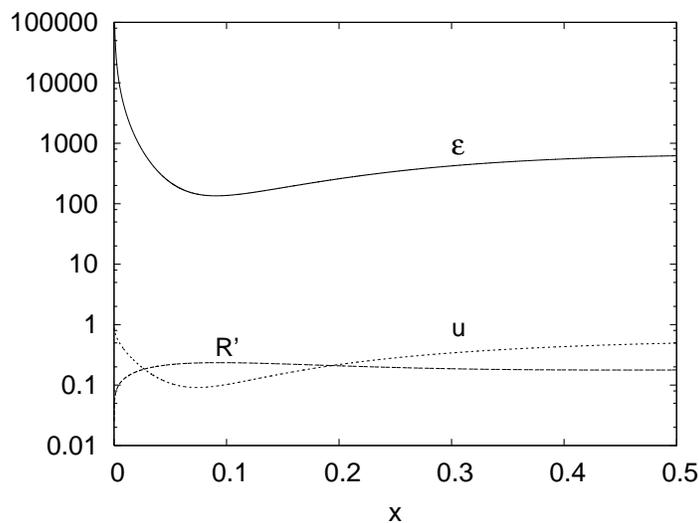}
 \caption{
 \label{uRprdeninter}
 \footnotesize
The functions $u(x)$, $R,_r(x)$ and $\epsilon(x)$ along the line $t = 0.05$ in
Fig. \ref{u=0}. }
 \end{center}
 \end{figure}

Fig. \ref{uRpratmidt} shows the same functions along the line $t = t_2$, where
$t_2$ lies between $t = 0.004$ and the right end of the the $u = 0$ contour in
Fig. \ref{u=0}; $t_2 = 0.025$. Finally, Fig. \ref{uRprdeninter} shows the
functions $u(x)$, $R,_r(x)$ and $\epsilon(x)$ along the line $t = t_3$, where
$t_3$ lies completely to the right of the $u = 0$ contour in Fig. \ref{u=0}. In
each case, the functions behave exactly in the way in which Fig. \ref{u=0}
implies they should.

\begin{figure}
 \begin{center}
 \includegraphics[scale = 0.7]{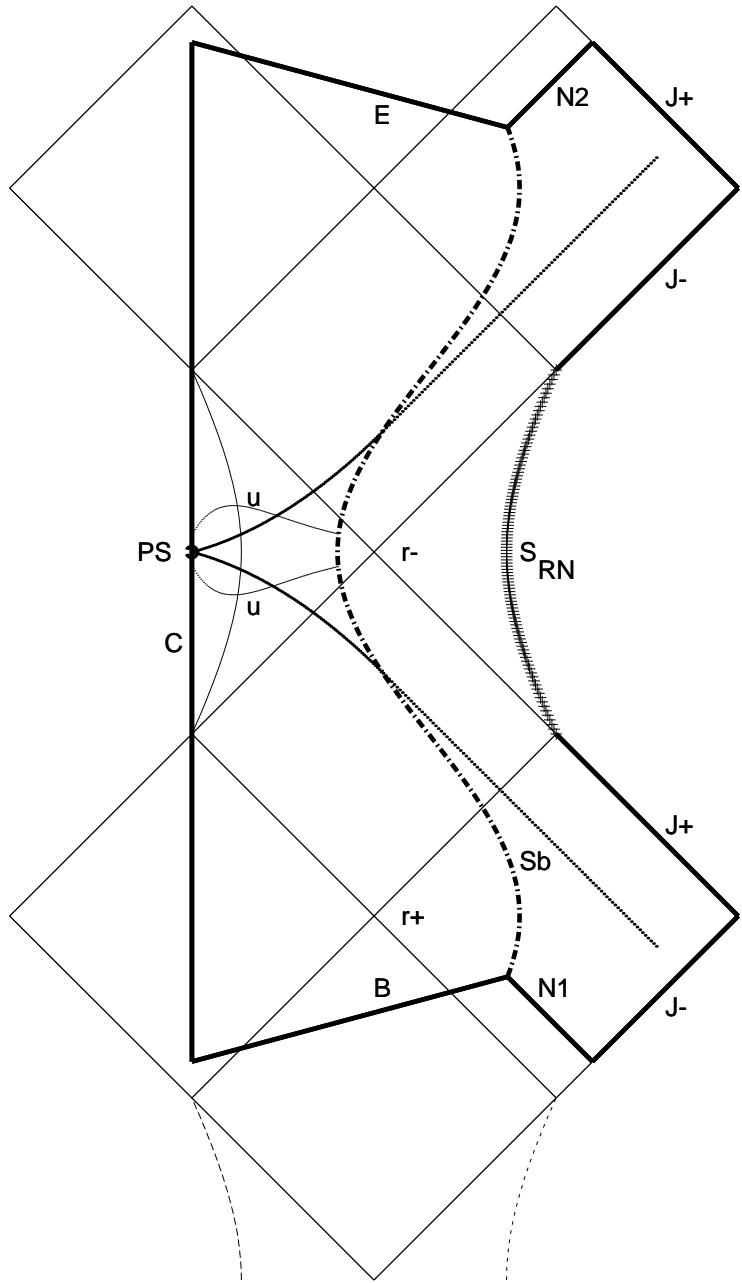}
 \caption{
 \label{chadumax2}
 \footnotesize
A schematic Penrose diagram for the configuration defined by eqs. (\ref{8.1}) --
(\ref{8.2}) and (\ref{8.5}). The image of the interior of the dust ball is
deformed to make its center $R = 0$ coincide, wherever possible, with the set $r
= 0$ of the background Reissner -- Nordstr\"{o}m diagram -- this is why the null
geodesics there are not straight lines at $45^{\circ}$. See more explanation in
the text. }
 \end{center}
 \end{figure}

Fig. 13 of Paper I, which showed the schematic Penrose diagram of the evolution
of our exemplary model, is still qualitatively correct. However, now it has to
be supplemented with a few new elements -- the images of the surfaces $u = 0$
and the light cone of the point singularity at the center. The complete version
of that diagram is shown in Fig. \ref{chadumax2}, and here is the explanation.

The diagram is written into the background of the Penrose diagram for the
maximally extended Reissner--Nordstr\"{o}m spacetime (thin lines). C is the
center of symmetry, Sb is the surface of the charged ball, S$_{\rm RN}$ is the
Reissner--Nordstr\"{o}m singularity. The interior of the body is encompassed by
the lines C, E, Sb and B; the only singularity that occurs within this area is
the point singularity at the center, marked PS. The dotted lines issuing from PS
mark the approximate position of the future and past light cone of the
singularity (i.e. the Cauchy horizon of the nonsingular part of the surface $R =
R_+$). Lines B and E connect the points in spacetime where the shell crossings
occur at different mass shells. N1 (N2) are the past- (future-) directed null
geodesics emanating from the points in which the shell crossings reach the
surface of the body (compare Fig. \ref{drawcycles2}). The line Sb should be
identified with the uppermost curve in Fig. \ref{drawcycles2}. The top end of Sb
is where the corresponding curve in Fig. \ref{drawcycles2} first crosses another
curve, the middle point of Sb is at $t = 0$ in Fig. \ref{drawcycles2}. The two
dotted arcs marked ``$u$'' symbolise the hypersurfaces $\epsilon = 0$, on which
the energy density changes from positive (below the lower curve) to negative and
then to positive again (above the upper curve). The distance between these arcs
is greatly exaggerated; at the proper scale of the figure they would seem to
coincide. The arcs are parts of the $u = 0$ contour shown in Fig. \ref{u=0}. In
Figs. \ref{drawcycles2} and \ref{chadumax2}, the surface of the body is at
smaller $x$ than the top of the $u = 0$ contour in Fig. \ref{u=0}. Every $N \neq
0$ world line of the dust necessarily enters the region of negative energy
density for a finite time interval around the instant of maximal compression.

\section{Permanent avoidance of singularity is impossible}\label{perm}

\setcounter{equation}{0}

In Paper I we hypothesised that a permanently nonsingular configuration of
charged dust might exist -- provided the period of oscillations is independent
of the mass ${\cal M}$. We show below that in our class of configurations the
period cannot be independent of ${\cal M}$ because conditions (1) -- (9)
prohibit this.

The condition $T_{c,{\cal M}} = 0$, if considered by explicitly differentiating
(\ref{2.29}), leads to a very complicated integral equation that we were not
able to handle. (The integral in (\ref{2.29}) cannot be explicitly calculated
without knowing the explicit form of $C({\cal M}, R)$.) However, let us recall
that both $T_p$ given by (\ref{2.28}) and $T_c$ given by (\ref{2.29}) are in
general functions of ${\cal M}$ only. Thus, we conclude that $T_c$ is a function
of $T_p$, and so the condition for the special case of $T_c$ being constant is
\begin{equation}\label{9.1}
\dr {T_c} {\cal M} = \dr {T_c} {T_p} \dr {T_p} {\cal M} = 0.
\end{equation}
Hence, $T_c$ will be constant when
\begin{equation}\label{9.2}
\dr {T_p} {\cal M} = 0 \Longleftrightarrow \dr {T_p} N = 0,
\end{equation}
since (\ref{2.21}) implies that $N,_{\cal M} \neq 0$. From (\ref{2.28}),
(\ref{2.9}) and $T_p = C =$ const we then find
\begin{equation}\label{9.3}
(- 2E)^{3/2} \equiv \left(1 - \Gamma^2\right)^{3/2} = C^{3/2} \left({\cal M} -
QQ,_N \Gamma\right).
\end{equation}
From here
\begin{equation}\label{9.4}
QQ,_N = - \frac 1 {\Gamma} \left(\frac {1 - \Gamma^2} C\right)^{3/2} + \frac
{\cal M} {\Gamma}.
\end{equation}
Given $\Gamma(N)$ and ${\cal M} (N)$ found from (\ref{2.21}), this determines
$Q(N)$ simply by integration.

We now verify whether (\ref{9.4}) is consistent with the regularity conditions
(\ref{2.22}) -- (\ref{2.25}) and the 9 conditions listed after (\ref{2.30}).

From (\ref{2.28}) we see that $T_p = C =$ const implies
\begin{equation}\label{9.5}
\lim_{{\cal M} \to 0} \frac M {\cal M} = \frac {D^{3/2} C} {2 \pi},
\end{equation}
where, as stated after (\ref{2.5}), $D$ may be zero. Then, comparing
(\ref{2.23}) with (\ref{2.27}), we see that
\begin{equation}\label{9.6}
\lim_{{\cal M} \to 0} \frac M {(-2 E) {\cal M}^{1/3}} = \alpha \neq 0, \qquad
\alpha < \infty,
\end{equation}
where $\alpha$ is a constant. This, together with (\ref{2.25}), means
\begin{equation}\label{9.7}
\lim_{{\cal M} \to 0} \frac M {D {\cal M}} = \alpha < \infty,
\end{equation}
i.e. that $D \neq 0$. But with $CD \neq 0$, eq. (\ref{9.5}) is in contradiction
with $\lim_{{\cal M} \to 0} M / {\cal M} \neq 0$ (see the remark at the end of
Sec. \ref{paper1}).

We have thus proven that a configuration with $E < 0$ cannot pulsate with the
period of pulsations being independent of the mass ${\cal M}$, while being
singularity-free for ever. Different periods for different mass shells will
necessarily cause shell crossings, during the second cycle at the latest (Fig.
\ref{drawcycles2} illustrates this).

Permanently nonsingular oscillations of weakly charged dust, with the period
independent of mass, are possible only when there is a central singularity. An
example of such a configuration results when $\Gamma = 1 / \cosh(bx)$, where $b$
is a constant.

A spherically symmetric charged dust configuration can be permanently
nonsingular only if it is static. Such configurations do indeed exist with
special forms of the arbitrary functions, as pointed out in Refs \cite{PlKr2006}
and \cite{KrBo2006}, but they are not interesting from the point of view of
avoiding singularities.

\section{Conclusions}\label{conclu}

The conclusion of this paper is: the weakly charged spherically symmetric dust
distribution considered here (${Q,_N}^2 < G / c^4$ at $N > 0$ and ${Q,_N}^2 = G
/ c^4$ at $N = 0$) must contain at least one of the following features:

1. A Big Bang/Big Crunch singularity;

2. A permanent central singularity;

3. A shell crossing singularity in a vicinity of the center

4. A finite time interval around the bounce instant in which the energy density
becomes negative, and a transient momentary singularity of infinite energy
density at a single point on the world line of the center of symmetry.

A fully nonsingular bounce is possible for a strongly charged configuration,
${Q,_N}^2 > G/c^4$, which can exist only with $E > 0$ -- see the comments after
(\ref{2.14}). This would be a collapse followed by a single bounce and
re-expansion to infinite size. This phenomenon occurs also in the Newtonian
limit -- the bounce here is caused by the prevalence of electrostatic repulsion
over gravitational attraction. Such an example was reportedly found, but never
published, by Ori and coworkers \cite{Oripriv}.

A permanently nonsingular pulsating configuration of spherically symmetric
charged dust does not exist. At most, a single full cycle of nonsingular bounce
can occur, and shell crossings will necessarily appear during the second
collapse phase. The nonsingular bounce occurs at $R > 0$, but the momentary
isolated singularity at the center of symmetry is still there.

The possibility of $\epsilon$ going negative in the presence of electric charges
does not seem to have been noticed and may need further work on its
interpretation. If the negative energy density region existed permanently in
some part of the space, with comoving boundary, then we might suspect that this
is a consequence of a bad choice of parameters that implies unphysical
properties in that region. However, here we have the case in which the energy
density is positive for some time, and then these same matter particles acquire
negative energy density in a time-interval around the bounce instant. This
suggests that there is some physical process involved in this, which should be
further investigated.

A question arises now. The uncharged limit of the family of configurations
defined by eqs. (\ref{2.1}) -- (\ref{2.10}), $Q = 0$, is the Lema\^{\i}tre --
Tolman (L--T) model \cite{Lema1933, Tolm1934}. For the latter, shell crossings
can be avoided, as is well-known since long ago \cite{HeLa1985}. Why, then, are
they unavoidable with $Q \neq 0$?

The answer is this: in the L--T model, the Big Bang (BB) or Big Crunch (BC) are
unavoidable (both are present when $E < 0$). In the cases that are colloquially
called ``free of shell crossings'', in reality the shell crossings are not
removed, but shifted to the epoch before the BB or after the BC, or both. Thus,
the shell crossings are no longer in the domain of physical applicability of the
model. When the BB/BC singularities are replaced with a smooth bounce in charged
dust, the shell crossings that were hidden on the other side of BB/BC become
physically accessible, and they terminate the evolution of the configuration.

\appendix

\setcounter{equation}{0}

\section{The proof that $R_+(N)$ coincides with $H_{\rm min}(N)$ at the local
extrema of $H_{\rm min}(N)$}\label{extrem}

We stated in the caption to Fig. 10 in Paper I, that the curves $H_{\rm min}(N)$
and $R_+(N)$ were not really tangent at the point where $H_{\rm min}(N)$ has its
maximum, but just close to one another. In fact, they not only were tangent at
that point, but {\it had to} be tangent, independently of the explicit forms of
the two functions. We show here that this is a general law: at every local
extremum of $H_{\rm min}(N)$ (call it $N_e$) we have $H_{\rm min}(N_e) =
R_+(N_e)$.

The radius of the inner R--N horizon as a function of $N$ is
\begin{equation}\label{a.1}
H_{\rm min}(N) = {\cal M} - \sqrt{{\cal M}^2 - GQ^2/c^4},
\end{equation}
where ${\cal M}$ and $Q$ are assumed to depend on $N$ through $x = N/N_0$. We do
not assume any explicit form of the functions ${\cal M}$ and $Q$, we only use
the general properties (\ref{2.9}), (\ref{2.15}) and (\ref{2.23}).

At every local extremum $N = N_e$ we have $H_{{\rm min},N} = 0$. Using
(\ref{2.23}), this means
\begin{equation}\label{a.2}
\left[\Gamma \sqrt{{\cal M}^2 - GQ^2/c^4}\right]_e = \left[{\cal M} \Gamma -
QQ,_N\right]_e,
\end{equation}
where the subscript $e$ denotes the value at the extremum. From here
\begin{equation}\label{a.3}
\left[\frac {GQ^2} {c^4}\right]_e = \left[\frac {QQ,_N} {\Gamma^2} \left(2 {\cal
M} \Gamma - QQ,_N\right)\right]_e.
\end{equation}
Substituting this value of $GQ^2/c^4$ in (\ref{a.1}) wet get
\begin{equation}\label{a.4}
H_{\rm min}(N_e) = QQ,_N/\Gamma.
\end{equation}
Substituting the same value of $GQ^2/c^4$ in (\ref{2.13}) for $R_+$ we find
\begin{equation}\label{a.5}
R_+(N_e) = QQ,_N/\Gamma = H_{\rm min}(N_e).
\end{equation}
$\square$

\setcounter{equation}{0}

\section{Proof of (\ref{3.2})}\label{negli}

For simplicity, to avoid physical coefficients, we introduce the functions
$p(x)$, $\mu(x)$ and $F_2(x)$ (where $x = N/N_0$, $N_0 =$ const) by
\begin{equation}\label{b.1}
Q(N) \df \pm \frac {\sqrt{G} N_0} {c^2}\ p(x), \qquad {\cal M} \df \frac {GN_0}
{c^4}\ \mu(x) \qquad M(N) \df \frac {GN_0} {c^4}\ F_2(x).
\end{equation}
Then, to prove (\ref{3.2}) we must prove that
\begin{equation}\label{b.2}
\lim_{x \to 0} \left[-2E p^2 \left({p,_x}^2 - 1\right)\right]/{F_2}^2 = 0.
\end{equation}

By condition (5), in a vicinity of $x = 0$, $p$ must be of the form
\begin{equation}\label{b.3}
p(x) = x + A_p x^{\alpha_p} + {\cal O}_{\alpha_p}, \Longrightarrow {p,_x}^2 - 1
= 2A_px^{\alpha_p - 1} + {\cal O}_{\alpha_p - 1},
\end{equation}
where $A_p$ and $\alpha_p > 1$ are constants, and ${\cal O}_{\alpha_p}$ is an
unspecified function with the property $\lim_{x \to 0} {\cal
O}_{\alpha_p}/x^{\alpha_p} = 0$. Then, by the regularity condition (\ref{2.24}),
\begin{equation}\label{b.4}
\Gamma = 1 + A_E x^{\alpha_E} + {\cal O}_{\alpha_E} \Longrightarrow -2E = -2 A_E
x^{\alpha_E} + {\cal O}_{\alpha_E},
\end{equation}
where $\alpha_E \geq 2/3$, by (\ref{2.25}). It follows from (\ref{2.21}) and
(\ref{2.8}) -- (\ref{2.9}) that
\begin{equation}\label{b.5}
\mu(x) = x + A_m x^{\alpha_E + 1} + {\cal O}_{\alpha_m}.
\end{equation}
Hence,
\begin{equation}\label{b.6}
F_2 \equiv \mu - pp,_x \Gamma = \left(A_m - A_E\right) x^{\alpha_E + 1} - A_p
\left(\alpha_p + 1\right)x^{\alpha_p} - {\cal O}_{\alpha_p} - {\cal O}_{\alpha_E
+ 1}.
\end{equation}
Thus $F_2$ is of order $\alpha_p$ or $\alpha_E + 1 \geq 5/3$, whichever limit is
lower. However, the limit via $\alpha_E$ leads to a central singularity, since
then, from (\ref{2.27}), $R/N^{1/3} \to 0$ as $N \to 0$, so (\ref{2.23}) will
not hold. Consequently, $F_2$ is of order $\alpha_p$, while the second term
under the square root in (\ref{3.1}), $-2E p^2 \left({p,_x}^2 - 1\right)$ is of
order $\alpha_2 = \alpha_E + \alpha_p + 1$. Suppose that $\alpha_2 \leq 2
\alpha_p$, so that the second term under the square root in (\ref{3.1}) is of
lower order than the first one. This translates to $\alpha_p \geq \alpha_E + 1$.
But condition (\ref{2.23}), in combination with (\ref{2.27}), requires that
$\alpha_p = \alpha_E + 1/3$, or else a central singularity will appear. Thus,
$-2E p^2 \left({p,_x}^2 - 1\right)$ must be of higher order in $x$ than
${F_2}^2$ in eq. (\ref{3.1}), i.e. (\ref{3.2}) must hold. $\square$

Now consider the case when we approach the center of symmetry along the locus of
the outer turning points, $R = R_-$. Then $u_- = \Gamma - QQ,_N / R_-$ remains
positive up to the center, as is seen from the reasoning below. The conclusion
that $-2E p^2 \left({p,_x}^2 - 1\right)$ is of higher order in $x$ than
${F_2}^2$ still applies, so we have
\begin{eqnarray}\label{b.7}
\lim_{N \to 0} u_- &\equiv& \lim_{N \to 0} \left[\Gamma + \frac {2EQQ,_N} {M +
\sqrt{M^2 - 2EQ^2 \left({Q,_N}^2 - G/c^4\right)}}\right] \nonumber \\
 &=& 1 + Q,_N(0) \lim_{N \to 0} EQ/M = 1 + Q,_N(0) \lim_{x \to 0} Ep/F_2.
\end{eqnarray}
We know from the above that $p(x) = x + A_p x^{\alpha_p} + {\cal O}_{\alpha_p}$,
$F_2 = - A_p \left(\alpha_p + 1\right)x^{\alpha_p} - {\cal O}_{\alpha_p}$ and $E
= A_E x^{\alpha_E} + {\cal O}_{\alpha_E}$, while the regularity condition
(\ref{2.23}) implies, via (\ref{2.27}), that $\alpha_E = \alpha_p - 1/3$. This,
taken together, implies that $ \lim_{x \to 0} Ep/F_2 = 0$, i.e. that $\lim_{N
\to 0} u_- = 1 > 0$, i.e. that $u_-$ remains positive up to the very center.

\section{Impossibility of a vacuole around $R = 0$ with $E < 0$}\label{novacu}

\setcounter{equation}{0}

Let us assume $E(x)$ as in (\ref{b.4}) and $p(x)$ in a form similar to
(\ref{b.3}):
\begin{equation}\label{c.1}
p(x) = \alpha x + A_p x^{\alpha_p} + {\cal O}_{\alpha_p}, \Longrightarrow p,_x =
\alpha + A_p \alpha_p x^{\alpha_p - 1} + {\cal O}_{\alpha_p - 1},
\end{equation}
For the function $\mu(x)$ this implies
\begin{equation}\label{c.2}
\mu = x + \frac {A_E} {\alpha_E + 1} x^{\alpha_E + 1} + {\cal O}_{\alpha_E + 1}.
\end{equation}
Then
\begin{eqnarray}\label{c.3}
F_2(x) &\equiv& \mu - p p,_x \Gamma = \left(1 - \alpha^2\right) x - A_E \left(2
\alpha^2 - \frac 1 {\alpha_E + 1}\right) x^{\alpha_E + 1} - \alpha A_p
\left(\alpha_p + 1\right) x^{\alpha_p} \nonumber \\
 &+& {\cal O}_{\alpha_p} + {\cal O}_{\alpha_E + 1}.
\end{eqnarray}
Now we have two cases: $\alpha^2 \neq 1$ and $\alpha^2 = 1$. In the first case
we have $F_2 = \left(1 - \alpha^2\right) x + {\cal O}_{1}(x)$. Thus, in order
that $M / E$ has a nonzero limit at $x \to 0$, it follows that $\alpha_E = 1$.
Then, if the terms $(M/2E)^2$ and $Q^2 \left({Q,_N}^2 - G / c^4\right)/(2E)$ in
(\ref{2.13}) are to be of the same order, the ratio $p^2 \left({p,_x}^2 -
1\right) / (2E)$ should have a nonzero finite limit at $x \to 0$. But, with
$\alpha^2 \neq 1$ and $\alpha_E = 1$ we have
\begin{eqnarray}\label{c.4}
&&\frac {p^2 \left({p,_x}^2 - 1\right)} {2E} = \frac {x^2 \left(\alpha + A_p
x^{\alpha_p - 1} + {\cal O}_{\alpha_p - 1}\right)^2 \left(\alpha^2 - 1 + 2 A_p
\alpha_p x^{\alpha_p - 1} + {\cal O}_{\alpha_p - 1}\right)} {2 A_E x + {\cal
O}_{1}} \nonumber \\
&& \llim{x \to 0} 0,
\end{eqnarray}
which has the consequence that $R_+ = 0$, i.e. the conditions $Q = {\cal M} = 0$
cannot be imposed at $R > 0$.

In the second case, $\alpha^2 = 1$, we can assume $\alpha = 1$ (nothing in the
equations depends on the sign of $Q$); then $F_2$ is of the order of either
$x^{\alpha_E + 1}$ or $x^{\alpha_p}$, whichever exponent is smaller. With
$\left(\alpha_E + 1\right)$ being smaller, the result is immediately seen: If
$F_2$ is of the order $x^{\alpha_E + 1}$ while $E$ is of the order
$x^{\alpha_E}$, then $\lim_{x \to 0} M/E = 0$, and $\lim_{x \to 0}R_+ = 0$. With
$\alpha_p$ being smaller we have
\begin{equation}\label{c.5}
\frac {p^2 \left({p,_x}^2 - 1\right)} {2E} = \frac {x^{\alpha_p + 1} \left(1 +
A_p x^{\alpha_p - 1} + {\cal O}_{\alpha_p - 1}\right)^2 \left(2 A_p \alpha_p +
{\cal O}_{0}\right)} {2 A_E x^{\alpha_E} + {\cal O}_{\alpha_E}}.
\end{equation}
This will have a finite limit at $x = 0$ if $\alpha_E = \alpha_p + 1$. But we
are considering the case when $F_2$ is of order $x^{\alpha_p}$, which means that
the limit of $F_2/E$, i.e. the limit of $M/E$ at $x \to 0$ will be infinite,
which again shows that the matching conditions cannot be fulfilled at a finite
nonzero $R$. $\square$

\bigskip

{\bf Acknowledgement} We are grateful to the referee for pointing out several
errors in an earlier version of this paper.

\end{document}